\documentclass[a4paper,11pt]{article}
  \pdfoutput=1
  \usepackage{silence}
  \usepackage{jheppub} 
  \usepackage{soul}

\usepackage{graphicx,epsfig}
\usepackage{amsmath}
\usepackage{amsfonts}
\usepackage{csquotes}
\usepackage{xcolor}

\usepackage{support-caption}
\usepackage{subcaption}
\usepackage[normalem]{ulem}

\newcommand{\He}{$^4$He}

\definecolor{darkgreen}{rgb}{0,0.5,0}

\usepackage{cleveref}

\begin{document}


\title{Signatures and Detection Prospects for sub-GeV Dark Matter with Superfluid Helium}

\author[a,b]{Yining You,}
\author[a]{Jordan Smolinsky,}
\author[a]{Wei Xue,} 
\author[a]{Konstantin Matchev,}
\author[a]{Keegan Gunther,} 
\author[a]{Yoonseok Lee,} 
\author[a]{Tarek Saab}

\affiliation[a]{Department of Physics, University of Florida, Gainesville, FL 32611, USA}
\affiliation[b]{Bard High School Early College DC, Washington, DC 20019, USA }

\abstract{We explore the possibility of using superfluid helium for direct detection of sub-GeV dark matter (DM). We discuss the relevant phenomenology resulting from the scattering of an incident dark matter particle on a Helium nucleus. Rather than directly exciting quasi-particles, DM in this mass range will interact with a single He atom, triggering an atomic cascade which eventually also includes emission and thermalization of quasi-particles. We present in detail the analytical framework needed for modeling these processes and determining the resulting flux of quasi-particles. We propose a novel method for detecting this flux with modern force-sensitive devices, such as nanoelectro-mechanical system (NEMS) oscillators, and derive the sensitivity projections for a generic sub-GeV DM detection experiment using such sensors.
}

\emailAdd{youy@ufl.edu}
\emailAdd{jsmolinsky@ufl.edu}
\emailAdd{weixue@ufl.edu}
\emailAdd{matchev@ufl.edu}
\emailAdd{kgunther@ufl.edu}
\emailAdd{ysl@ufl.edu}
\emailAdd{tsaab@ufl.edu}

\maketitle

\section{Introduction}

Although the existence of dark matter (DM) has been definitively supported by gravitational evidence \cite{Zwicky:1933gu,Bertone:2016nfn,Arbey:2021gdg}, 
its nature is still one of the biggest mysteries in modern physics. 
Recent theoretical developments \cite{Knapen:2017xzo,lin2019tasi} and tighter exclusion limits from both direct detection experiments \cite{Agnese_2019, Armengaud_2019, Aprile_2019, Aprile_2018, Alkhatib_2021, Agnes_2018, Abdelhameed_2019} and the Large Hadron Collider (LHC) at CERN \cite{Behr:2022tyz} are increasingly pointing towards DM being lighter than first thought, with mass less than $\sim$1 GeV. This motivates the design of new direct detection experiments which specifically target the sub-GeV and sub-MeV DM mass range \cite{alexander2016dark,battaglieri2017us,Essig:2022dfa}.

Direct detection experiments achieve the greatest sensitivity when the fundamental excitation of the target material corresponds to the recoil energy or momentum scale from the scattering dark matter. Therefore, superfluid $^4$He, with a spectrum of quasi-particle excitations with momenta $\lesssim {\rm keV}$ \cite{PhysRevB.103.104516,landau1987statistical}, becomes an excellent target material candidate for detecting very light dark matter \cite{Hertel:2018aal,Knapen:2016cue,Schutz:2016tid,Acanfora_2019,Baym:2020uos,Caputo:2019cyg,Caputo:2019xum,Caputo:2019ywq,Caputo:2020sys}.
Although superfluid $^4$He has been considered as a target for neutrino and dark matter studies since the 1980s \cite{Lanou:1987eq,Huang:2007jh,bandler1993projected,Bradley:1996cu,Winkelmann:2006pw,Winkelmann:2006rg,Lanou:1988iq,Bandler:1991ep,Bandler:1992zz,Adams:1996ge}, its experimental application developed swiftly in recent years as the particle physics community refocused its interest on the search for light dark matter candidates \cite{alexander2016dark,battaglieri2017us,lin2019tasi,Bottino:2002ry,Bottino:2003cz,Shelton:2010ta,Feng:2008ya,Foot:2008nw,CMS:2012lmn,Fermi-LAT:2011vow}. One effort in applying superfluid $^4$He is the search for light Weakly Interacting Massive Particles (WIMPs) of mass below 10 GeV \cite{Guo:2013dt,Ito:2013cqa,Osterman:2020xkb}. Due to the kinematics, the minimum velocity of light WIMPs required for elastic nuclear recoil is fairly low for Helium compared to heavier materials like Xenon and Germanium, and this fact could offer some additional sensitivity beyond that of the Xenon experiment. Other efforts have focused on the search for light dark matter candidates with masses below MeV \cite{Schutz:2016tid,Knapen:2016cue,Acanfora_2019,Caputo:2019cyg,Baym:2020uos}. They utilize the fact that the fundamental excitation (quasi-particle) spectrum of the superfluid $^4$He matches the recoil energy scale or momentum scale of the scattering dark matter. Therefore, each dark matter scattering event leads to the production of one or two phonon quasi-particles in the superfluid, for which the event rates and cross sections can be derived analytically. However, the practical detection of single excitations in the superfluid remains an experimental challenge.

In lieu of these previous proposals, in a previous theoretical work \cite{Matchev:2021fuw} we studied the possibility to use superfluid $^4$He to detect DM in the {\it sub-GeV} mass range, i.e., for a range of DM masses between 1 MeV and 1 GeV.\footnote{For a calculation of the multi-phonon production in a crystal target over the whole keV to GeV DM mass range, see \cite{Campbell-Deem:2022fqm}.} 
The DM of our Milky way local group has a typical velocity of magnitude $10^{-3}c$ \cite{Kavanagh:2016xfi,Lee:2012pf,Radick:2020qip,Necib:2018igl,Ibarra:2017mzt}. Therefore, a DM particle with mass above 1 MeV has a de Broglie wavelength smaller than $\sim {\cal O}(1) \textup{~\AA}$, the distance between helium atoms in the superfluid. As a result, the DM initially scatters with a single helium atom rather than producing multiple phonon quasi-particles. As we shall elaborate, this mass range conveniently produces a predominantly neutral atomic cascade, which has the advantages of (a) increasing the yield of quasi-particles and improving the sensitivity compared to that in the sub-MeV range; and (b) simplifying the modeling and projections of the experimental reach.

\begin{figure}[tb]
\centering
  \includegraphics[width=1\columnwidth]{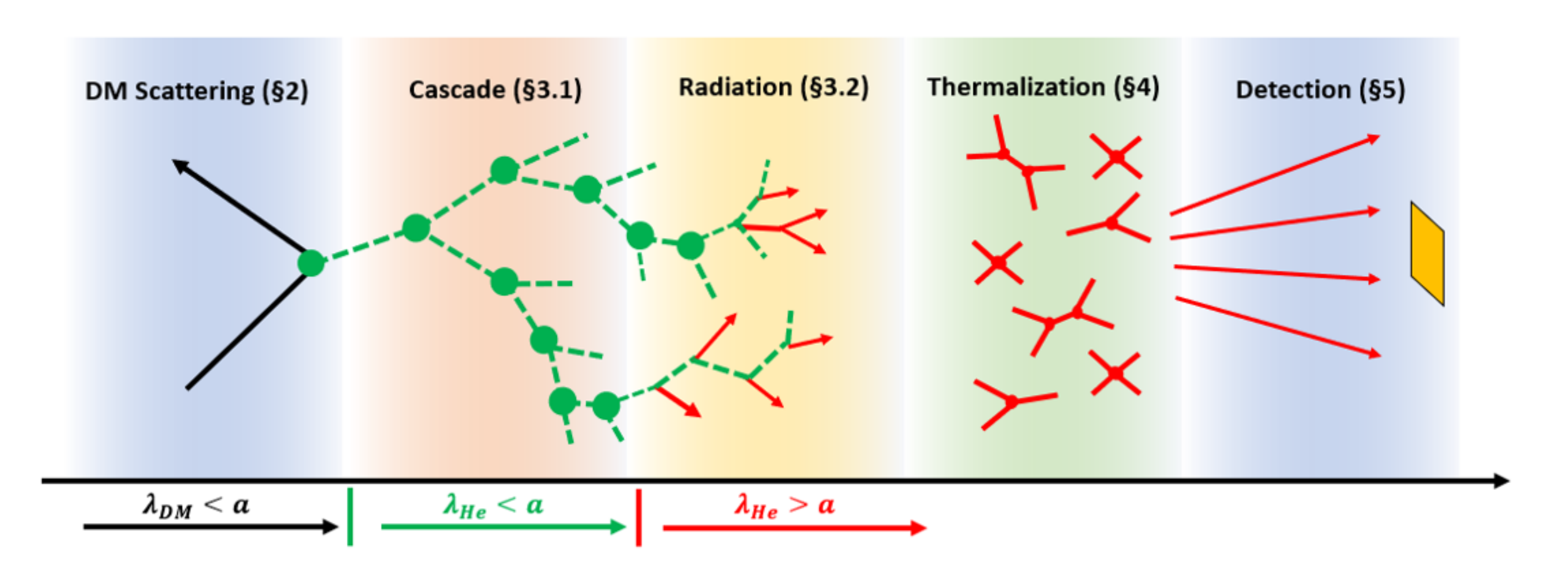}
\caption{Schematic illustration of a dark matter particle with mass between 1 MeV and 1 GeV scattering off superfluid \He. The de Broglie wavelength $\lambda_{DM}$ of the incoming DM particle is smaller than the inter-atomic spacing $a$, so that it scatters with a helium atom at rest (green dot). The dashed green lines correspond to scattered helium atoms freely propagating until being scattered by other helium atoms (green dots). This cascade process becomes subdominant when the de Broglie wavelengths $\lambda_{He}$ of the scattered helium atoms are comparable or larger than the atomic spacing, i.e., when $\lambda_{He} > a$. Quasi-particles emitted by the He atoms are plotted with red lines. The quasi-particles rapidly interact with each other via various channels (depicted by the Feynman diagrams). Finally, a flux of quasi-particles with a Bose-Einstein distribution reaches the detector (yellow square).}
\label{fig:schematicplot}
\end{figure}

The basic physics processes are illustrated in the pentaptych of figure~\ref{fig:schematicplot}. In the first panel, the recoiling helium atom inherits the $\sim$ MeV momentum scale of the DM. It will proceed and scatter against other helium atoms, which themselves will scatter in turn, and so on, producing an avalanche of atoms (second panel). At the end of this atomic cascade, the initial recoil momentum has been distributed over a large number of daughter atoms moving isotropically with respect to the initial scattering location. These atoms have a de Broglie wavelength comparable or larger than the inter-atomic distance of $\sim {\cal O}(1) \textup{~\AA}$. At this level, the 2-2 elastic scattering becomes subdominant and each low momentum atom interacts with the superfluid background and radiates quasi-particle excitations (including but not limited to phonons), as shown in the third panel. In our previous theoretical work, we focused on constructing an effective field theory (EFT) to describe this process \cite{Matchev:2021fuw}. Here we build a numerical simulation which explicitly models the (relevant processes in the) atomic cascade. We will find that the quasi-particles are produced in a small region $\sim {\cal O}(1)$ nm around the DM impact, and subsequently thermalize (fourth panel). Using the thermal distribution of quasi-particles we can trace the transport of momentum through the superfluid and derive the momentum imparted to a generic nanoelectro-mechanical system (NEMS) sensor suspended within (fifth panel). This allows us to derive the sensitivity projections for a generic sub-GeV DM detection experiment using a NEMS sensor.

The organization of our paper is as follows. In \cref{sec:DMscatterHe}, we review the elastic scattering between a DM particle and an initial helium atom. The calculation produces the helium recoil rate profiles for DM of different masses. In \cref{sec:otherprocesses}, we review other scattering processes that may happen in the superfluid, and explain that they are subdominant for the sub-GeV regime. In \cref{sec:HeCascade}, we review the well known quantum mechanics treatment of helium-helium atomic cascade. In \cref{sec:HeEmitsQP}, we review our previous theory proposal's result on the Lagrangian construction of helium atoms emitting quasi-particles. The calculation provides a quasi-particle production rate for a helium atom of an arbitrary momentum. In \cref{sec:IntofQPs}, we review the parameters of quasi-particles and show that they are thermalized by the time all quasi-particles are radiated from the slow helium atoms. In \cref{sec:Thermalization}, we determine the temperature of the thermalized quasi-particle system by two different methods --- an analytical approximation or a Monte Carlo simulation. In \cref{sec:flux&force}, we calculate the momentum signal on the oscillator sensor in a realistic spatial configuration. Finally, we present the experimental constraints on the DM-Nucleus coupling strength. The five sections are chronologically ordered, with each section feeding a particle profile to the next section (see \cref{fig:schematicplot}). In \cref{sec:conclude}, we conclude our findings.

\section{\bf Dark matter scattering off Helium atoms}
\label{sec:DMscatterHe}

\subsection{DM elastic scattering}

We begin by considering the DM scattering rate off helium atoms via elastic nuclear recoil, which is the dominant process in the sub-GeV DM mass range. As shown in \cref{fig:schematicplot}, the superfluid Helium response to a dark matter scattering event depends only on the momentum of the initial recoiling He atom and is agnostic to the microscopic physics of the dark matter sector. For the purposes of our analysis, in this section we shall introduce a simple toy model of dark matter interactions which will be used to derive experimental sensitivity limits in \cref{sec:flux&force}. Suppose that a fermionic dark matter particle $\chi$ couples to the nucleon $n$ (we assume equal couplings to neutrons and protons) via a heavy scalar mediator particle $\phi$,
\begin{equation}
    \mathcal{L}_\text{int}=g_\chi \, \phi\, \bar{\chi}\chi+g_n\, \phi \, \bar{n}n.
\end{equation}
The differential DM-Helium cross section is
\begin{equation}
\frac{ {\rm d} \, \sigma_{\chi\rm{\,He}}}{ {\rm d} {\bf q}^2}
= \frac{ A^2  g_\chi^2 g_n^2} {4 \pi v^2} \frac{F_n^2(\textbf{q}^2) } {(\textbf{q}^2+m_\phi^2)^2}
\simeq 
\frac{ A^2  g_\chi^2 g_n^2} { 4\pi  v^2} \frac{1 } {m_\phi^4} \, ,
\label{eq:DM-He-Sigma}
\end{equation}
where $A=4$ is the atomic number of Helium. To arrive at the last result, we have taken the massive mediator limit $m_\phi\gg q_{\max}$ and approximated the nuclear form factor as $F(\textbf{q}^2)\to 1$, which is justified by the fact that the inverse momentum of a sub-GeV mass DM, 
$\hbar/q$, is much larger than the nucleus size $\sim$ fm.

The differential nuclear recoil rate ${\rm d}\Gamma / {\rm d} q^2$
is linearly proportional to the number of target helium atoms $N_{\rm He}$, the local DM density $\rho_\odot=0.3$ GeV/cm$^3$, 
and depends on the DM velocity distribution $f_\chi(\textbf{v})$ and the differential cross section (\cref{eq:DM-He-Sigma}), 
\begin{eqnarray}
      \frac{dR}{d\textbf{q}^2} &=& \frac{\rho_\odot}{m_\chi} N_\text{He} \, \int d^3v f_\chi(v) v \, \frac{d\sigma_{\chi\text{He}}}{d\textbf{q}^2}
        \nonumber   \\ 
        &=& \frac{\rho_\odot}{m_\chi} N_\text{He} \, \int d^3v f_\chi(v) \,   \frac{A^2 \, \sigma_{\chi n} }{4 v  \mu_{\chi n}^2 }  \ ,
   \label{eq:differentialrate}
\end{eqnarray}
where in the second line 
we have introduced the DM-nucleon reduced mass, $ \mu_{\chi n}$, and the
DM-nucleon cross section $\sigma_{\chi n} = \frac{g_\chi^2 g_n^2 \mu_{\chi n}^2}{ \pi m_\phi^4}$. 
The DM velocity distribution $f_\chi(\textbf{v})$ in our local frame is taken as the Maxwell-Boltzmann distribution with an escape velocity cutoff $v_{\text{esc}}$,
\begin{equation}
    f_{\chi}(\mathbf{v})=\frac{\Theta(v_{\text{esc}}-|\mathbf{v}+\mathbf{v}_{e}|)}{N(v_{0},v_{\text{esc}})}\exp\left(-\frac{(\mathbf{v}+\mathbf{v}_{e})^{2}}{v_{0}^{2}}\right),
    \label{eq:MB distribution}
\end{equation}
where the normalization factor
  $$  N(v_{0},v_{\text{esc}})=\pi^{3/2}v_{0}^{3}
  \left[\text{erf}\left(\frac{v_{\text{esc}}}{v_{0}}\right)-2\frac{v_{\text{esc}}}{v_{0}}\exp\left({-\frac{v_{\text{esc}}^2}{v_{0}^2}}\right)\right]$$
ensures that $\int d^3 v f(v) = 1$.
The Heaviside function in \cref{eq:MB distribution} constraints the DM velocity in the galactic frame to be below the escape velocity. Here we choose 
the velocity dispersion $v_{0} = 220\ \text{km/s}$,
the velocity of the earth $v_{e} = 240\ \text{km/s}$, and the escape velocity $v_{esc} = 500\, \text{km/s}$. 

\begin{figure}[tb]
\centering
  \includegraphics[width=0.8\columnwidth]{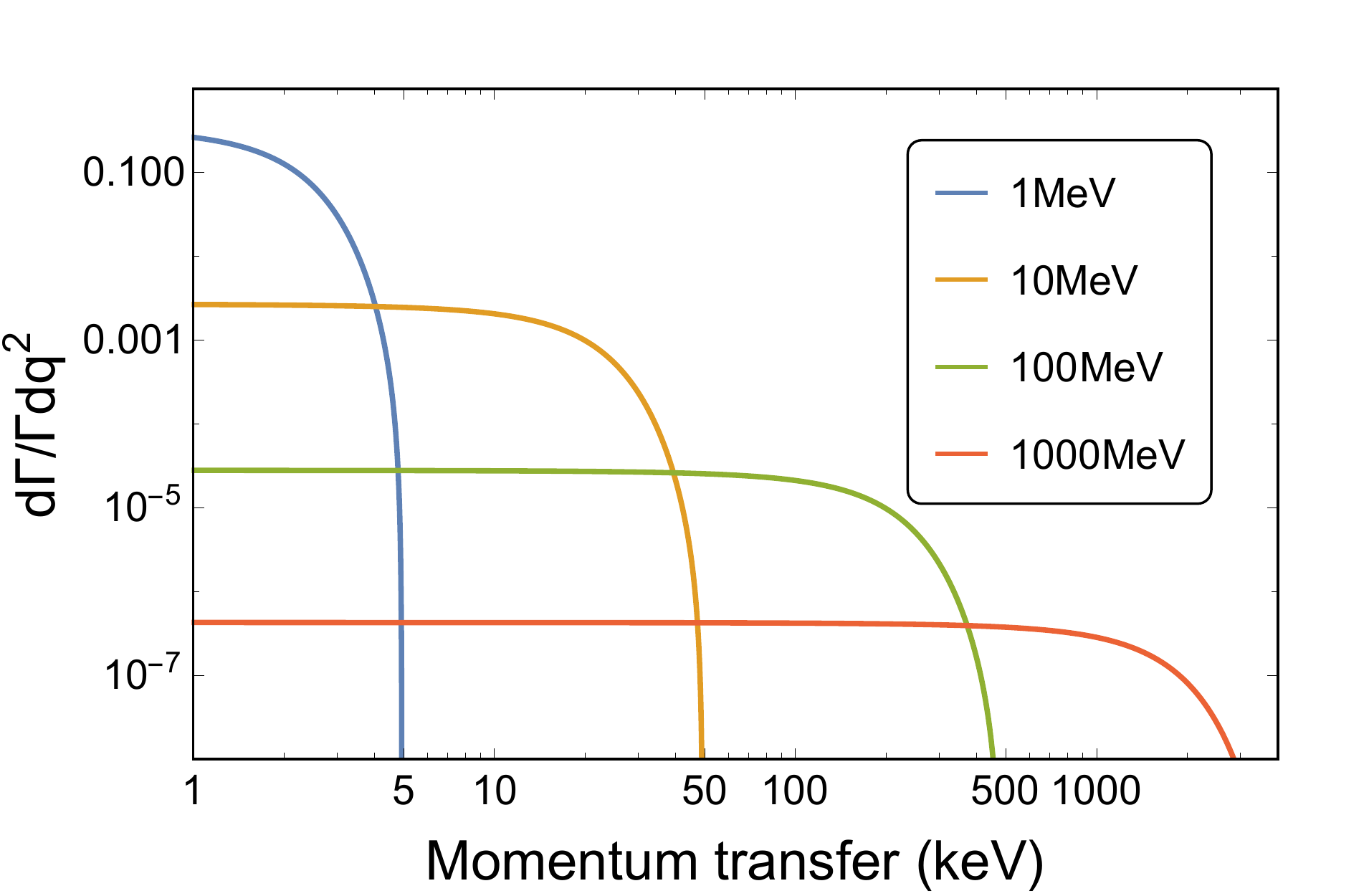}
\caption{Differential recoil rate for DM-Helium scattering $\frac{d\Gamma }{\Gamma dq^2}$. Results are shown for several values of the DM mass: $m_\chi=1$ MeV (blue), $m_\chi=10$ MeV (orange), $m_\chi=100$ MeV (green) and $m_\chi=1000$ MeV (red).
}
\label{fig:DM-Helium recoil}
\end{figure}

The recoil momentum of the Helium atom is the essential quantity which governs the spectrum of the produced quasi-particles. By using eqns.~(\ref{eq:DM-He-Sigma}-\ref{eq:MB distribution}), we obtain the differential Helium recoil rate shown in \cref{fig:DM-Helium recoil}. We note that heavier dark matter particles imply harder recoil spectra.

\subsection{Other DM scattering signatures: direct production of quasi-particles and inelastic scattering} 
\label{sec:otherprocesses}

In addition to elastic scattering off individual helium atoms, there are several other DM-induced processes, which are subdominant in the MeV-GeV DM mass range targeted in this paper, but may become dominant for DM masses either below MeV or above GeV. In the sub-MeV mass range, the DM de Broglie wavelength is larger than several$\textup{~\AA}$, spanning several helium atoms in the liquid. 
Such a light DM particle could directly produce quasi-particle excitations via
coherent scatterings.
At higher masses above a GeV, the exchange momentum is large enough to trigger excitation and ionization of helium.

\paragraph{Coherent quasi-particle production.}
The general scattering rate involving both coherent and incoherent quasi-particle production can be derived using the many-body quantization method \cite{Matchev:2021fuw,Knapen:2016cue}
\begin{equation}
\frac{dR}{d\textbf{q}^2\, d\omega} = \frac{\rho_\odot}{m_\chi} N_\text{He} \, \int d^3v f(v) \,   \frac{A^2 \, \sigma_{\chi n} }{4 v  \mu_{\chi n}^2 }\, S(\textbf{q},\omega)  \ ,
\label{eq:directQP}
\end{equation}
where the form factor $S(\textbf{q},\omega)$ is the Dynamic Structure Function (DSF) of the superfluid. At low momentum, $q \sim {\rm \rm keV}$, the DSF is measured from the neutron energy loss in neutron scattering experiments \cite{Silver:1989zz}. At high momentum, $q \gg {\rm keV}$, the scattering is
mainly incoherent, which is initiated by nuclear recoil and produces quasi-particles by the subsequent helium radiation. 
The integration of the incoherent part of the DSF $\int\,d\omega S_{inc}(\textbf{q},\omega) = 1$ \cite{griffin1993excitations}, 
so that the quasi-particle production rate is identical to the nuclear recoil rate in eq.~(\ref{eq:differentialrate}), as expected.
In principle, the DSF $S(\textbf{q},\omega)$ can be applied to study DM scattering at any momentum, but 
at high momentum $S(\textbf{q},\omega)$ is unknown and it is practical to use the nuclear recoil formula eq.~(\ref{eq:differentialrate}). The coherent scattering becomes dominant at low momentum, where the measured or simulated DSF \cite{Silver:1989zz,campbell2015dynamic} could be employed 
to evaluate the relevant quasi-particle production rate. The production rate of phonon quasi-particles can also be derived
using either the impurity method  \cite{landau1941theory,Matchev:2021fuw} or effective field theory \cite{Acanfora_2019,Nicolis_2018,PhysRevLett.119.260402}.

\begin{figure}[tb]
\centering
  \includegraphics[width=0.8\columnwidth]{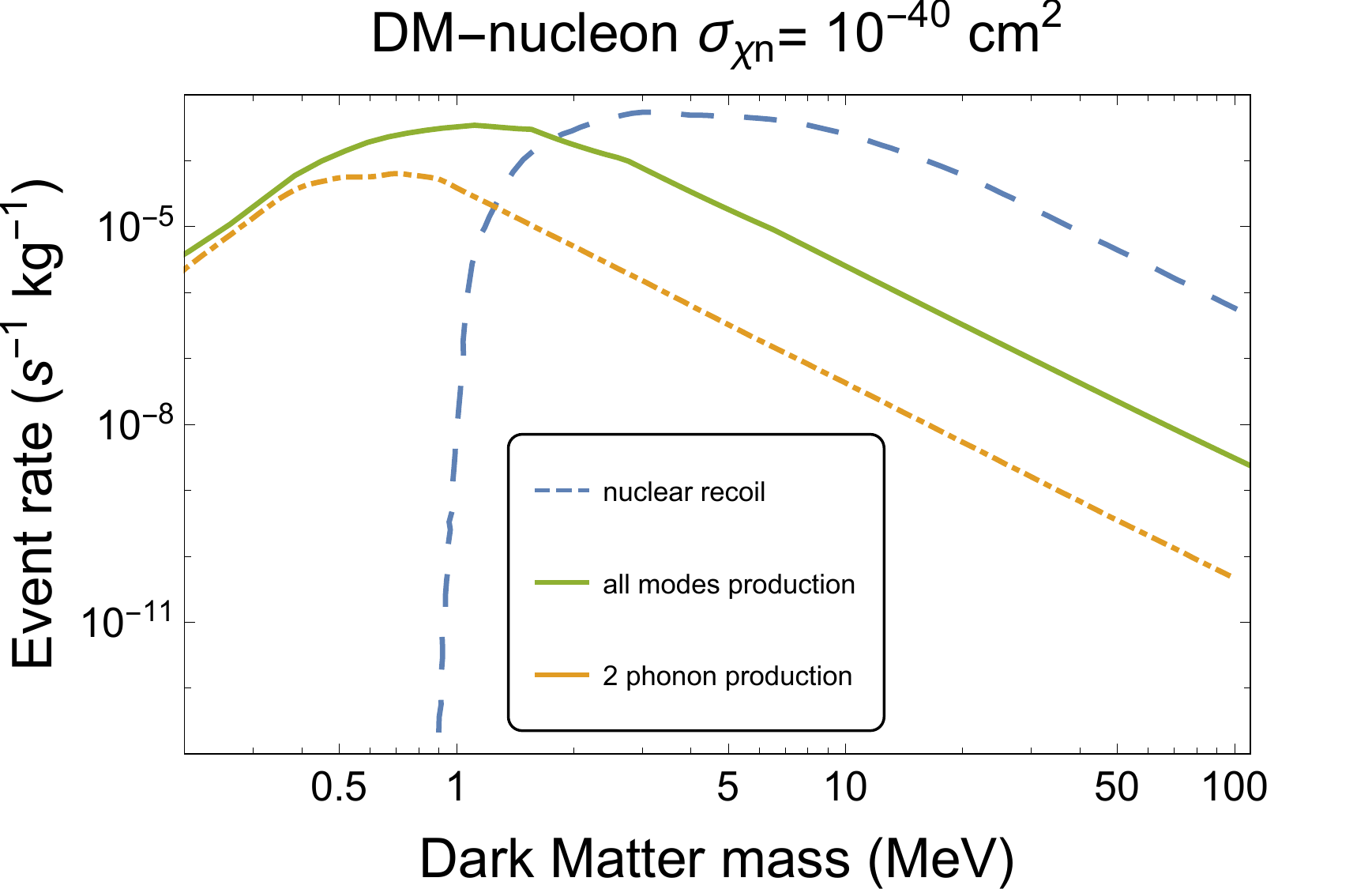}
\caption{Event rates for the processes of nuclear elastic scattering (blue dashed line), coherent quasi-particle emission (green solid line), and double phonon emission (orange dot-dashed line). Here we take a dark matter-nucleon cross section of $10^{-40}$ cm$^2$ and 1 kg of target superfluid. The solid green line represents the event rate of single and multiple quasi-particle production with a net energy threshold of 1.2 meV. The nuclear recoil assumes no nuclear recoil below 4.5 keV. The dot-dashed orange data is taken from \cite{Acanfora_2019}, where the 2-phonon detection threshold is at an energy of 1 meV. For a discussion of these threshold values, see refs.~\cite{Acanfora_2019,Knapen:2016cue}.}
\label{fig:DMtoPhonon}
\end{figure}

To compare elastic scattering with coherent emission of quasi-particles, we show their respective event rates in \cref{fig:DMtoPhonon}. The nuclear recoil rate (blue dashed line) is dominant for masses above $\sim 1$ MeV, while coherent emission (green solid line) dominates for sub-MeV mass DM. The nuclear scattering rate is derived from eq.~(\ref{eq:differentialrate}) by integrating over the momentum ${\bf q}^2$ starting from $q_{\rm c} = 4.5 \, {\rm keV}$, assuming that below $q_{c}$ DM will scatter coherently with the superfluid. 
The coherent scattering rate is derived from \cref{eq:directQP} by integrating over ${\bf q}^2$ up to $q_{c}=4.5$ keV, with the DSF data given in \cite{campbell2015dynamic}. The DSF $S(\textbf{q},\omega)$ peaks at $ \omega \sim {\rm meV}$ and decreases quickly at high energy, because producing a large
number of quasi-particles from a single coherent scattering is suppressed by couplings and phase space. The exchanged momentum in coherent scattering is restricted up to $q_c$, while the momentum of nuclear recoil is $q \simeq m_\chi v $. 
Then for masses larger than MeV the coherent rate scales as $1/m_\chi^3$, while the incoherent rate goes only like $1/m_\chi$ and is therefore dominant at high mass. Furthermore, in each event, the coherent scattering generates ${\cal O}(1)$ 
quasi-particles, but the incoherent scattering could produce many more quasi-particles, depending on the DM mass.

\begin{figure}[tb]
\centering
  \includegraphics[width=0.8\columnwidth]{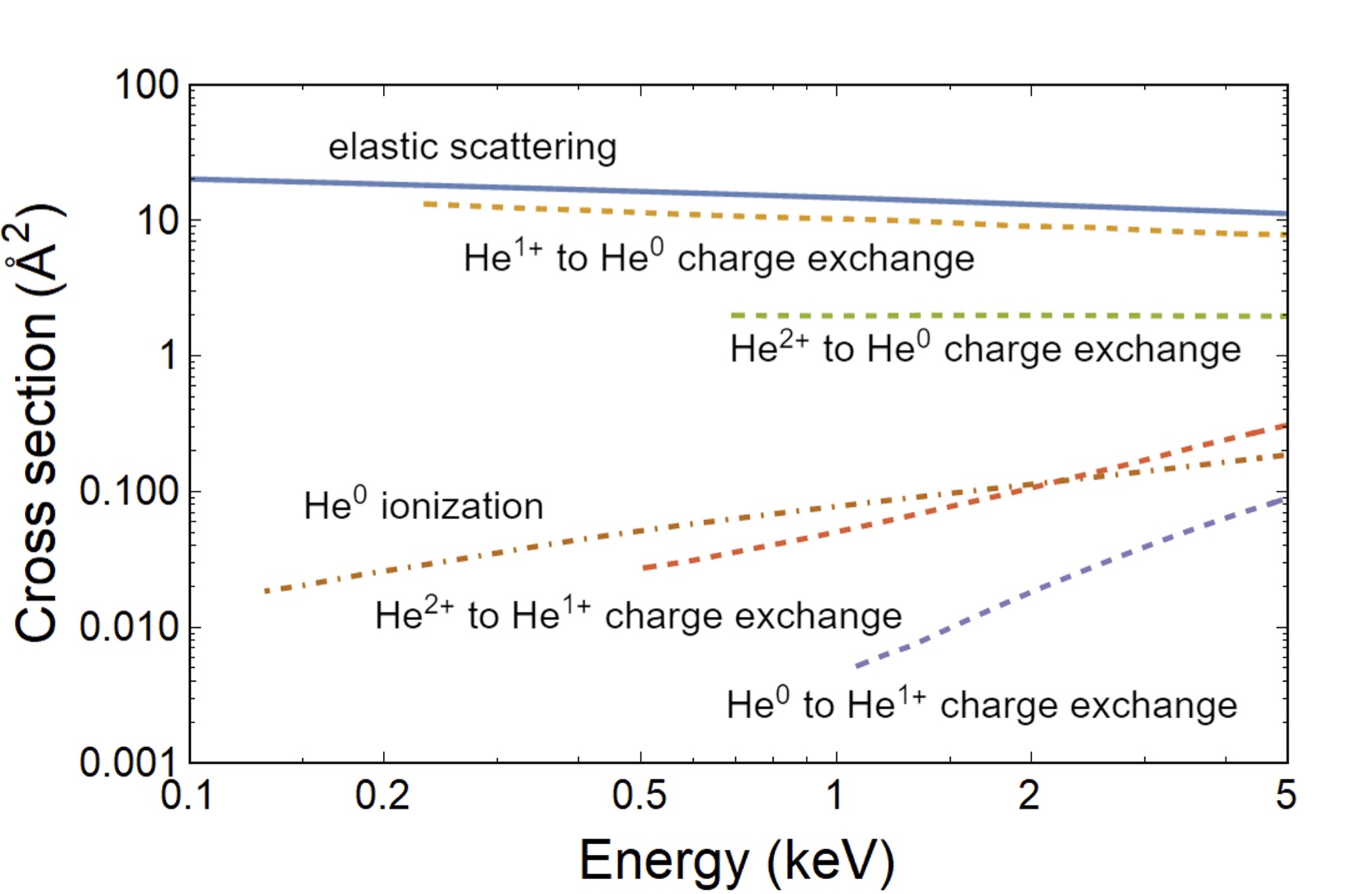}
\caption{A summary plot of helium atom elastic scattering, helium ion charge exchange scattering, and helium ion ionization cross sections \cite{Guo:2013dt}, which interpolate the results in \cite{PhysRev.38.1342,PhysRevA.9.2434,cramer1957elastic,Hegerberg_1978,PhysRevA.39.4440,pivovar1962electron,PhysRevA.32.829,PhysRevA.20.1816,Grozdanov_1980,Fulton_1966,hertel1964cross,Stich_1985,RevModPhys.30.1137,WU198857,Hvelplund_1976,doi:10.1246/bcsj.49.933,thomas_1927,PhysRev.124.128,PhysRev.135.A1575,PhysRev.109.355,PhysRev.178.271,PhysRevA.36.2585,PhysRevA.63.062717,Shah_1985,osti_4196582,Heer1965ExcitationOH,Mei:2007jn}.
The curves represent helium elastic scattering (blue solid), He$^{1+}$ to He$^{0}$ (He$^{2+}$ to He$^{0}$, He$^{2+}$ to He$^{1+}$, He$^{0}$ to He$^{1+}$) charge exchange scattering (dashed lines), and He$^{0}$ ionization (brown dot-dashed). The cross sections of 
other possible channels are lower than $10^{-3}\textup{~\AA}^2$.}
\label{fig:Charge exchange + Ionization}
\end{figure}

\paragraph{Charge exchange, ionization and excitation processes.} 

In the mass range above $\sim 0.1$ GeV, the DM kinetic energy and helium nuclear recoil energy are larger than $100$ eV, which is sufficient to excite or ionize a helium atom. Considering the interactions among helium atoms,  
the elastic scatterings are still predominant and neutral helium atoms
constitute the majority of the final products after these processes. 
This is illustrated in \cref{fig:Charge exchange + Ionization}, showing the leading cross sections of 
charge exchange processes and ionization processes.
After the primary helium atom or ion are produced, they continue interacting with the helium atoms in the superfluid through the processes of charge exchange scattering \cite{PhysRev.109.355,PhysRev.38.1342,PhysRevA.9.2434,cramer1957elastic,Hegerberg_1978,PhysRevA.39.4440,pivovar1962electron,PhysRevA.32.829,PhysRevA.20.1816,Grozdanov_1980,Fulton_1966,hertel1964cross,Stich_1985,RevModPhys.30.1137,WU198857,Hvelplund_1976}, ionization \cite{doi:10.1246/bcsj.49.933,thomas_1927,PhysRev.124.128,PhysRev.135.A1575,PhysRev.109.355,PhysRev.178.271,PhysRevA.36.2585,PhysRevA.63.062717,Shah_1985}, and excitation \cite{osti_4196582,Heer1965ExcitationOH, PhysRev.124.128,Mei:2007jn}.  
At the energy scale below $\sim$ keV, the cross sections for neutral helium producing helium ions are much smaller than the reverse processes. 
As a result, neutral helium atoms are the dominant final products once these interactions reach equilibrium \cite{Guo:2013dt}.

 The recoil energy is dissipated in the following channels: 
 (1) ionized atom conversion into neutral atoms, 
 (2) neutral atom cascade, (3) quasi-particle production by atoms, (4) decay of excited atoms into IR photon and singlet and triplet dimer excimers $\text{A}^1 \Sigma_u^+$ and $\text{a}^3 \Sigma_u^+$. Existing efforts in the literature \cite{Guo:2013dt,Hertel:2018aal,Ito:2011cy,Ito:2013cqa,Adams:1995mk} have led to a clear estimation of the partition of recoil energy among these interaction channels. Essentially, quasi-particle production is the only relevant process at recoil energies below $20$ eV, and it accounts for the dominant fraction all the way up to $100$ keV.

\section{Helium atom cascade} 
\label{sec:HeCascade}

In this section we will discuss the theoretical description and the details of our simulation of the neutral atomic cascade. 
We simplify the DM-helium scattering model by considering only the neutral atoms portion of the cascade.
This is justified by the numerical analysis in \cite{Guo:2013dt} and the summary in \cref{sec:DMscatterHe} 
showing that the primary recoiling helium atom/ion will quickly produce a neutral atomic cascade.

\subsection{Elastic scattering of Helium atoms}
\label{sec:elastic}

A recoiling helium atom with momentum in the keV to MeV range has de Broglie wavelength smaller than the inter-atomic distance in the superfluid. Therefore, the initial scattered atom triggers a cascade, i.e., a series of $2\to2$ scatterings among helium atoms within the fluid. Because the kinetic energy is comparable to the helium atomic potential \cite{bennewitz1972he,feltgen1973determination,bishop1977low}, the $2\to2$ atomic scattering at this energy scale is non-perturbative. Nonetheless, given a potential function, we may numerically solve the Schrodinger equation using Partial Wave Expansion. Consider the wavefunction $\Psi(\textbf{r})$ as a function of the displacement vector $\textbf{r}$ between two helium atoms. The initial and final asymptotic boundary conditions constrain the wavefunction as follows:
\begin{equation}
\Psi(\boldsymbol{r})|_{r\to\infty}=\exp(ikz)+f_\Psi(\theta,k)\frac{\exp(ikr)}{r}.
\label{eq:helium wave}
\end{equation}
The first term is an initial plane wave, the second term is an outgoing spherical wave weighted with a scattering amplitude $f_\Psi(\theta,k)$, and $k$ is the reduced momentum in the center-of-mass frame (equal to one half of the incoming atom momentum in the lab frame).

Expanding \cref{eq:helium wave} in terms of Legendre Polynomials $P_l$, the Schrodinger equation of the system can be solved numerically, and the scattering amplitude $f_\Psi(\theta,k)$ depends on the phase shift $\delta_l(k)$, where $l$ is the angular index of the Legendre expansion. In \cite{Matchev:2021fuw} we calculated the scattering cross section
\begin{equation}
\sigma_{\text{He-He}}(k)=\frac{8\pi}{k^{2}}\sum_{l\in\text{even}}(2l+1)\sin^{2}\delta_{l}(k),
\label{eq:HeCrossSection}
\end{equation}
using phase variation \cite{calogero1967variable} and WKB approximation \cite{miller1969wkb,miller1971additional}. We use the total scattering rate for the estimation in the next section of the helium cascade time which is related to the final shower size and quasi-particle number density. The differential cross section is incorporated in the simulation of the atomic cascade, and has the form
\begin{equation}
\frac{d\sigma_{\text{He-He}}}{d\Omega}=\frac{1}{k^2}\left|\sum_{l\in\text{even}}(2l+1)\left[e^{2i\delta_l(k)}-1\right]P_l(\cos\theta)\right|^2.
\label{eq:HeCSDiff}
\end{equation}
For the convenience of comparison with the next section,  in \cref{fig:HeRad} we plot the experimental total cross section using data from \cite{feltgen1973determination}  (red dashed curve).

\subsection{Quasi-particle emission from Helium atoms}
\label{sec:HeEmitsQP}

The elastic scatterings keep dissipating the energy to more and more helium atoms, and decreasing their momenta.
When a helium atom's momentum drops below $10$ keV, the atom's de Broglie wavelength becomes comparable to the inter-atomic distance of the superfluid. The moving atom then collectively scatters against  the surrounding atoms in the fluid and produces quasi-particles. 

Unlike the case of sub-MeV dark matter directly producing quasi-particles \cite{Schutz:2016tid,Knapen:2016cue,Acanfora_2019,Caputo:2019cyg}, the helium emission
process is non-perturbative. In \cite{Matchev:2021fuw} we proposed an effective $U(1)$ current-current coupling between a helium atom and the superfluid:
\begin{equation}
\mathcal{L}_{JJ}=\lambda_{1}\frac{1}{m_{{\rm He}}\Lambda}J^{0}J_{{\rm He}}^{0}+\lambda_{2}\frac{m_{{\rm He}}}{\Lambda^{3}}J^{i}J_{{\rm He}}^{i} \, .\label{eq:HeJJ}
\end{equation}
$\Lambda$ is the cutoff scale of the superfluid EFT. We can estimate $\Lambda$ using the inverse of the atomic space, $\Lambda \sim {\cal O}(1) \, {\rm keV}$. 
For the phonon, the cutoff is related to 
the energy density of the superfluid $\rho$ and the sound speed of phonon,  $\Lambda = (\rho \, c_s)^{1/4} \simeq 0.83 \, {\rm keV}$. $J^0$ is the number density operator of the superfluid, and its normalized matrix element is the Dynamic Structure Function (DSF) \cite{Baym:2020uos,Knapen:2016cue,silver1988theory}. The general form of the total emission rate of multiple quasi-particles is as follows:
\vspace{2mm}
\begin{equation}
\Gamma_\text{inel} =  \frac{2\pi\rho}{m_\text{He}} \int \frac{d^{3}k} {(2\pi)^{3}}\left(\frac{\lambda_1}{m_{\text{He}}\Lambda}
+ \frac{\lambda_2 m_{\text{He}}}{\Lambda^3} \frac{\boldsymbol{v}_{\text{He}}\cdot\boldsymbol{k}\omega}{k^2}\right)^{2}S(\boldsymbol{k},\omega),
\label{eq:multi-particle radiation}
\end{equation}
where $\rho$ is the superfluid mass density, $v_\text{He}$ is the initial helium velocity, and $S(\textbf{k},\omega)$ is the DSF. 

\begin{figure}[tb]
\centering
\includegraphics[width=0.8\columnwidth]{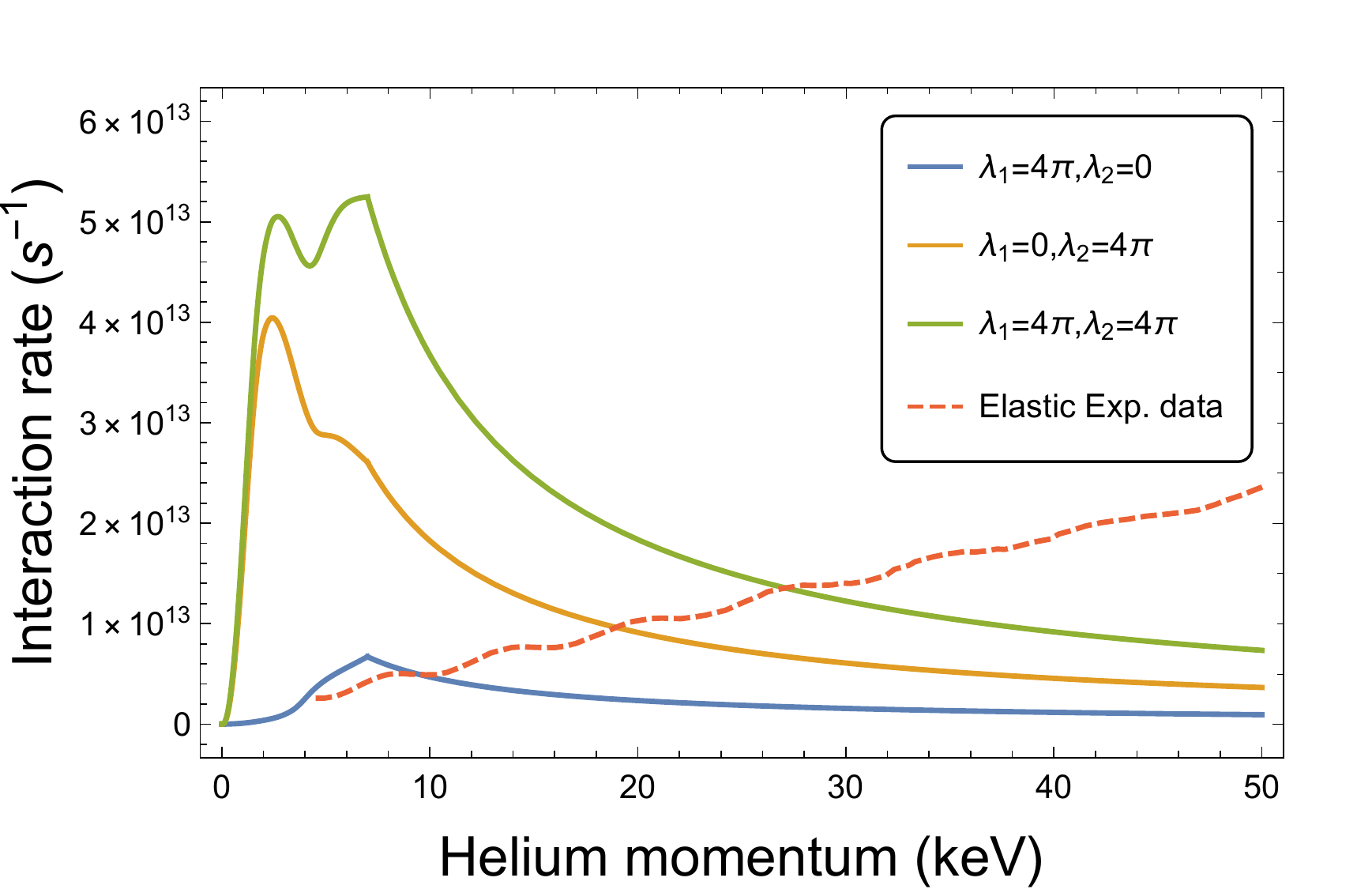}
\caption{
The rate of helium emitting one quasi-particle (green, orange, and blue lines) and the experimental rate of elastic atomic scattering (red dashed line). The helium atom momentum is the lab frame momentum, consistent in all curves. The elastic atomic scattering rate is converted from the experimental scattering cross section given in \cite{feltgen1973determination}. We test three combinations for the coupling parameters in \cref{eq:multi-particle radiation}, namely, $\lambda_1=4\pi,\lambda_2=0$ (blue), $\lambda_1=0,\lambda_2=4\pi$ (orange), and $\lambda_1=4\pi,\lambda_2=4\pi$ (green). The orange and green lines match the expectation that at low momentum helium predominantly radiates quasi-particles.
}
\label{fig:HeRad}
\end{figure}
For the purpose of comparing the elastic atomic scattering cross-sections with that for quasi-particle emission, in \cref{fig:HeRad} we show the emission rate for {\em single} quasi-particles. This is because our numerical integration of the phase space for multiple production $\sim\int d^3k\,S(k,\omega)$ and of the phase space for single production $\sim\int d^3k\, S(k)\,\delta (\omega-\omega(k))$ with current available data shows that single quasi-particle production is dominant. In \cref{fig:HeRad} we show results for several combinations of the coupling parameters: $\lambda_1=4\pi,\lambda_2=0$ (blue), $\lambda_1=0,\lambda_2=4\pi$ (orange), and $\lambda_1=4\pi,\lambda_2=4\pi$ (green).
Although the values of $\lambda_1$ and $\lambda_2$ are unknown, in this region the system is strongly coupled, so we take the value of $4\pi$. 
In the simulation, we choose $\lambda_1=4\pi,\lambda_2=0$, but when we consider the energy/momentum conservation and thermalization after the quasi-particle production, 
the final quasi-particle flux will not be sensitive to the specific couplings due to the thermalization. 
The physical expectation of low energy scattering originates from the two requirements: (1) helium atoms predominantly radiate quasi-particles rather than elastically scatter with each other in the superfluid; (2) quasi-particle modes only exist below $\sim 7$ keV. The first requirement shows that the vector coupling in \cref{eq:HeJJ} must exist. The second requirement restricts the momentum integration to an upper cutoff $\sim 7$ keV. Below the cutoff, the emission rate increases with the momentum of the atom because of the phase space enhancement. Beyond the cutoff, the emission rate decreases because the phase space integration in \cref{eq:multi-particle radiation} has reached its maximum, leaving powers of helium atom momentum only in the denominator of the result. The curves in \cref{fig:HeRad} thus have a cusp at $7$ keV because of this different behavior below and above the cutoff.

\subsection{Monte Carlo simulation of the Helium cascade and radiation of quasiparticles}
\label{sec:MCsimulation}

In this section, we will present our simulation results about the developing helium cascade and the radiation of quasi-particles from fast moving helium atoms. We do not include the subsequent quasi-particle decays and quasi-particle self-interactions, whose treatment is postponed for the next section.

Because of the large number of daughter particles which must necessarily be produced to conserve both energy and momentum, little information about the overall final structure of the shower may be gleaned from a direct analytical treatment. However, because the individual interactions of the shower constituents are quantum mechanical and probabilistic, the problem is amenable to a Monte Carlo approach. In particular the distance that each helium atom travels between interactions, the type of interaction it experiences, and the subsequent evolution of its daughter particles are all properly described by probability distributions (all necessary results were derived and/or collected in \cite{Matchev:2021fuw}), which we exploit to generate ensembles of simulated events.  

The momentum-dependent cross section $\sigma_\text{el}$ of {\em elastic} atomic helium scattering is known from experiment \cite{feltgen1973determination} (see red line in \cref{fig:HeRad}). On the other hand, the rate of {\em inelastic} emission of quasi-particles has been computed in \cite{Matchev:2021fuw} and is given by \cref{eq:multi-particle radiation}. This rate can be cast as an inelastic ``cross section'' according to the heuristic
\begin{equation}
    \sigma_\text{inel} = \frac{\Gamma_\text{inel}}{n_\text{He} v} \ ,
\end{equation}
where $n_\text{He}=\rho/m_\text{He}$ denotes the number density of helium atoms in the superfluid. 
The total cross section of these processes defines a mean free path
\begin{equation}
    \ell_0 = \frac{1}{n_\text{He} (\sigma_\text{inel} + \sigma_\text{el})}
\end{equation}
and thus a probability distribution 
\begin{equation}
    P_\text{interaction}(\ell) \propto \exp\left[-\frac{\ell}{\ell_0}\right]
\end{equation}
that describes the distance $\ell$ an energetic helium atom is expected to travel before interacting with the superfluid to produce either another energetic helium atom or a quasi-particle. The respective probabilities for each type of daughter particle are given by
\begin{equation}
    P_\text{el} = \frac{\sigma_\text{el}}{\sigma_\text{el} + \sigma_\text{inel}} \ , \hspace{0.5in} P_\text{inel} = \frac{\sigma_\text{inel}}{\sigma_\text{el} + \sigma_\text{inel}} \ .
\end{equation}
Note that the cross sections and consequently the length scale of the probability distribution are all functions of the helium momentum.
At this point we have all the ingredients needed to describe the structure of the Monte Carlo simulation: each simulated event begins as a single helium atom at the origin with its momentum oriented along the $z$-axis. Using this momentum we evaluate the cross sections of both processes and sample the resulting distribution to determine what type of interaction the helium atom experiences and where this interaction occurs. The momenta of daughter particles are sampled from the appropriate differential rates provided in sections~\ref{sec:elastic} and \ref{sec:HeEmitsQP} above, and the cross sections of those daughter particles are evaluated anew. In this way the simulation proceeds recursively generating new daughter particles and tracking their trajectories between collisions. This process is depicted schematically in the second and third panels of \cref{fig:schematicplot}.

\begin{figure}[hptb]
\centering
\includegraphics[width=0.31\columnwidth]{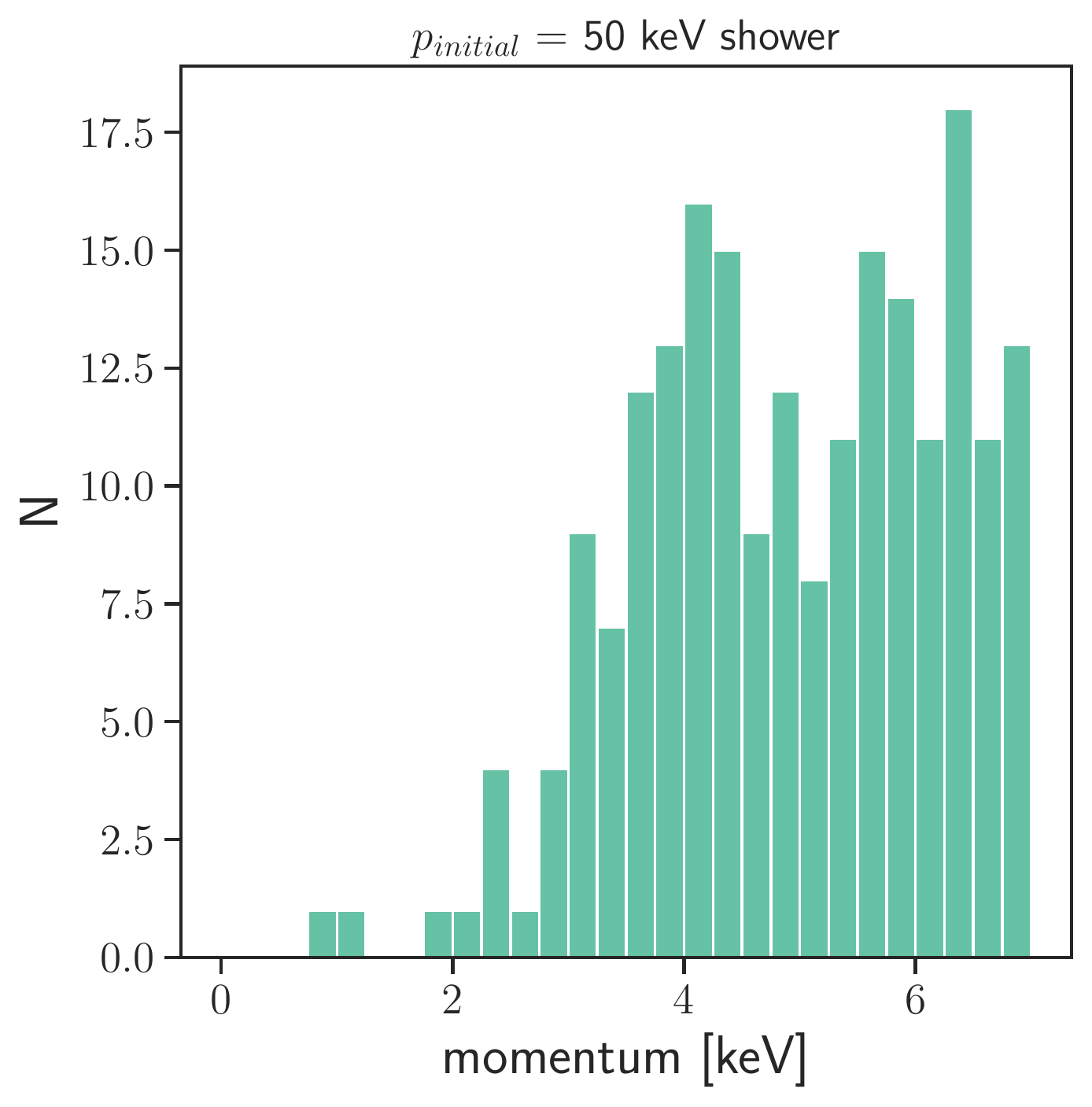}~~~~~~
\includegraphics[width=0.31\columnwidth]{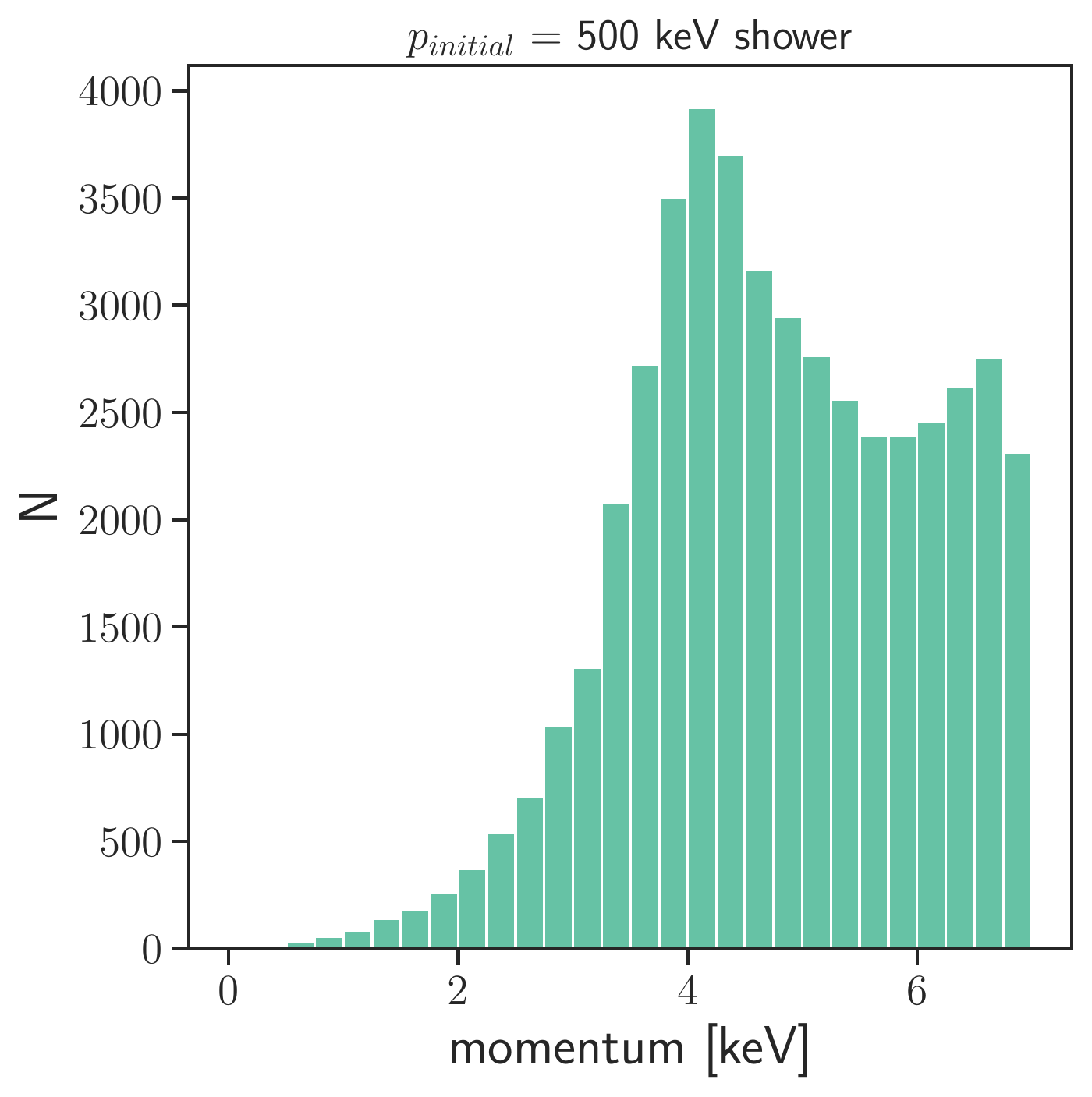}\\
\includegraphics[width=0.31\columnwidth]{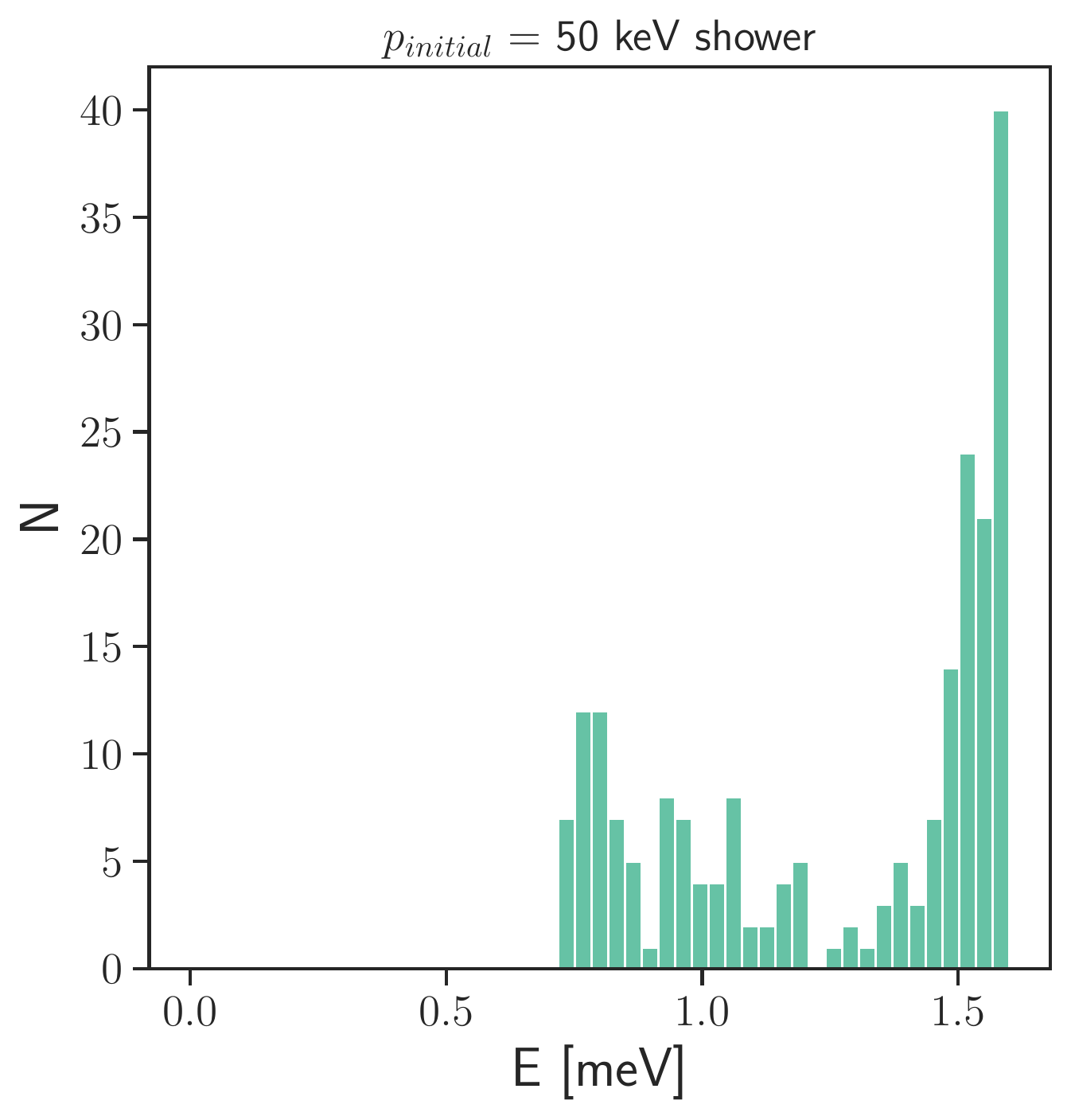}~~~~~~
\includegraphics[width=0.31\columnwidth]{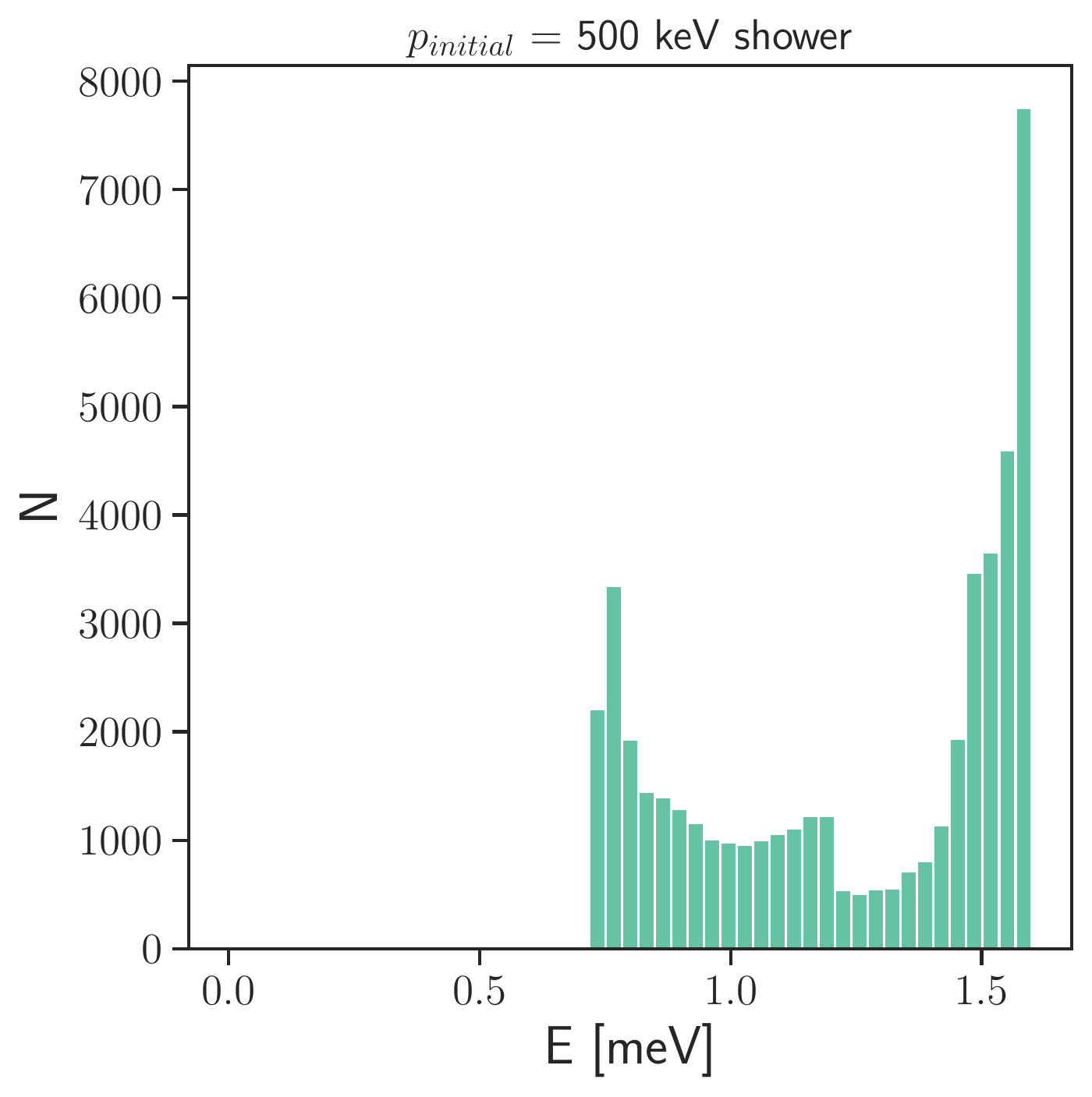}\\
\includegraphics[width=0.31\columnwidth]{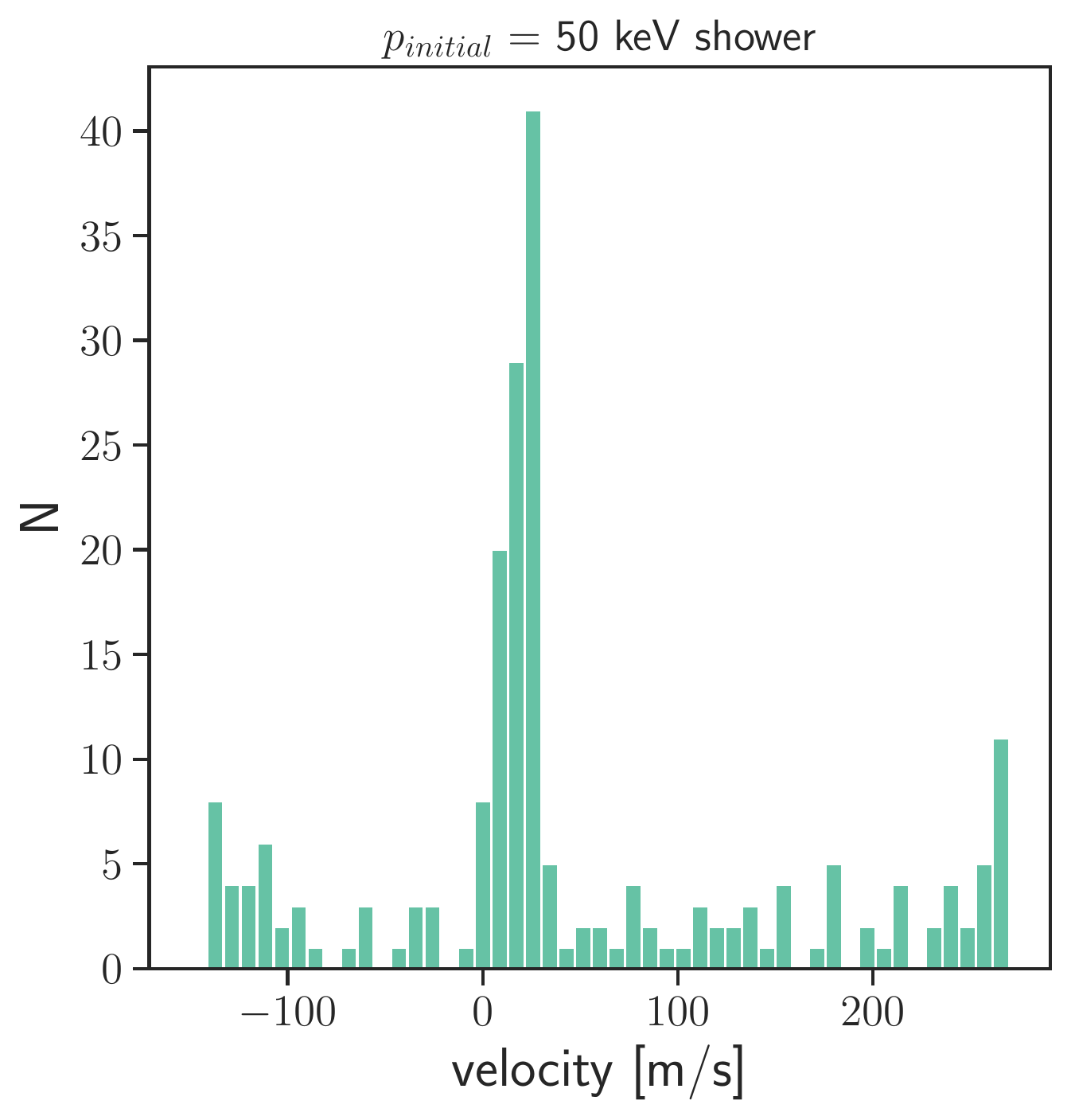}~~~~~~
\includegraphics[width=0.31\columnwidth]{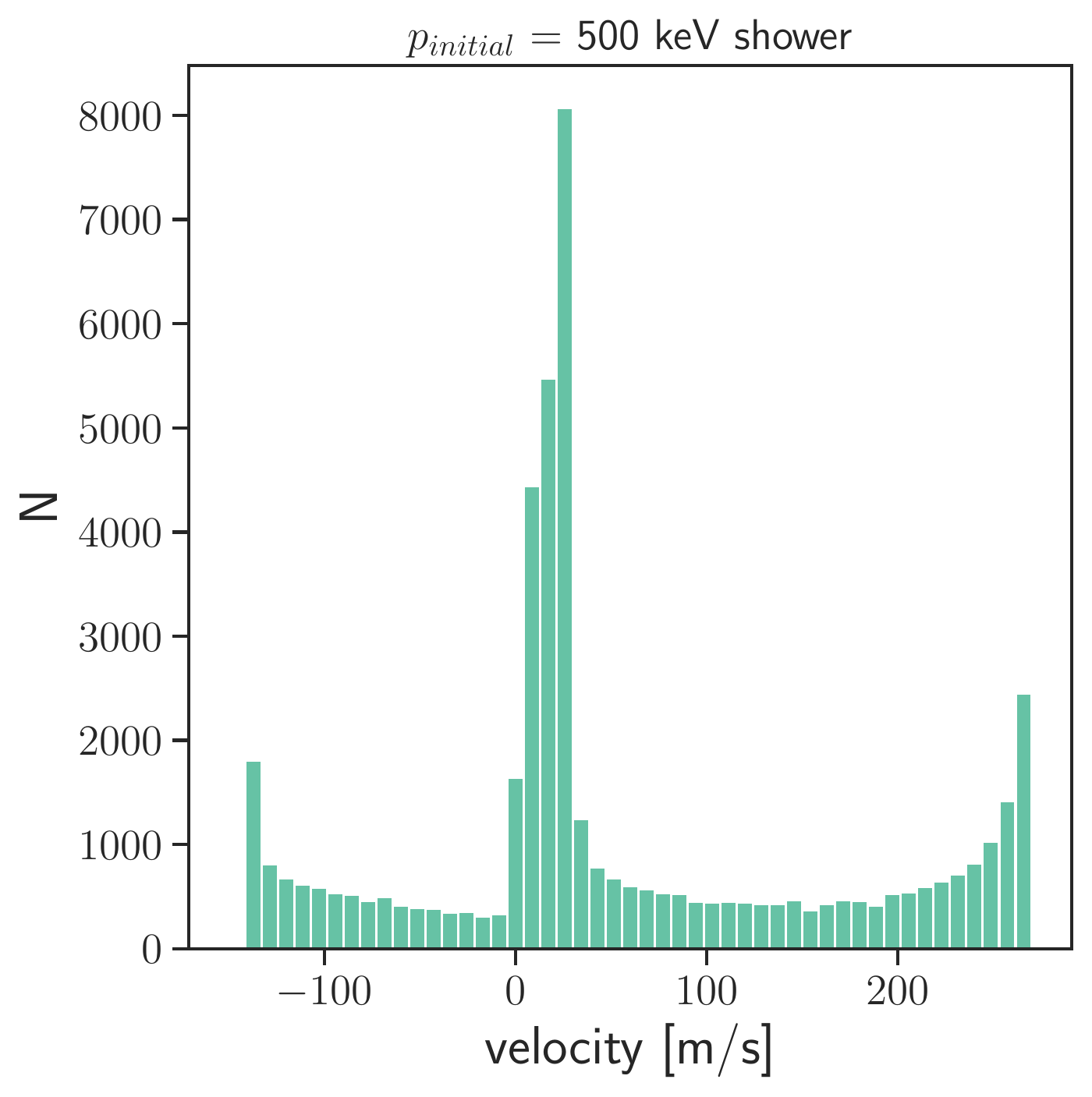}\\
\includegraphics[width=0.31\columnwidth]{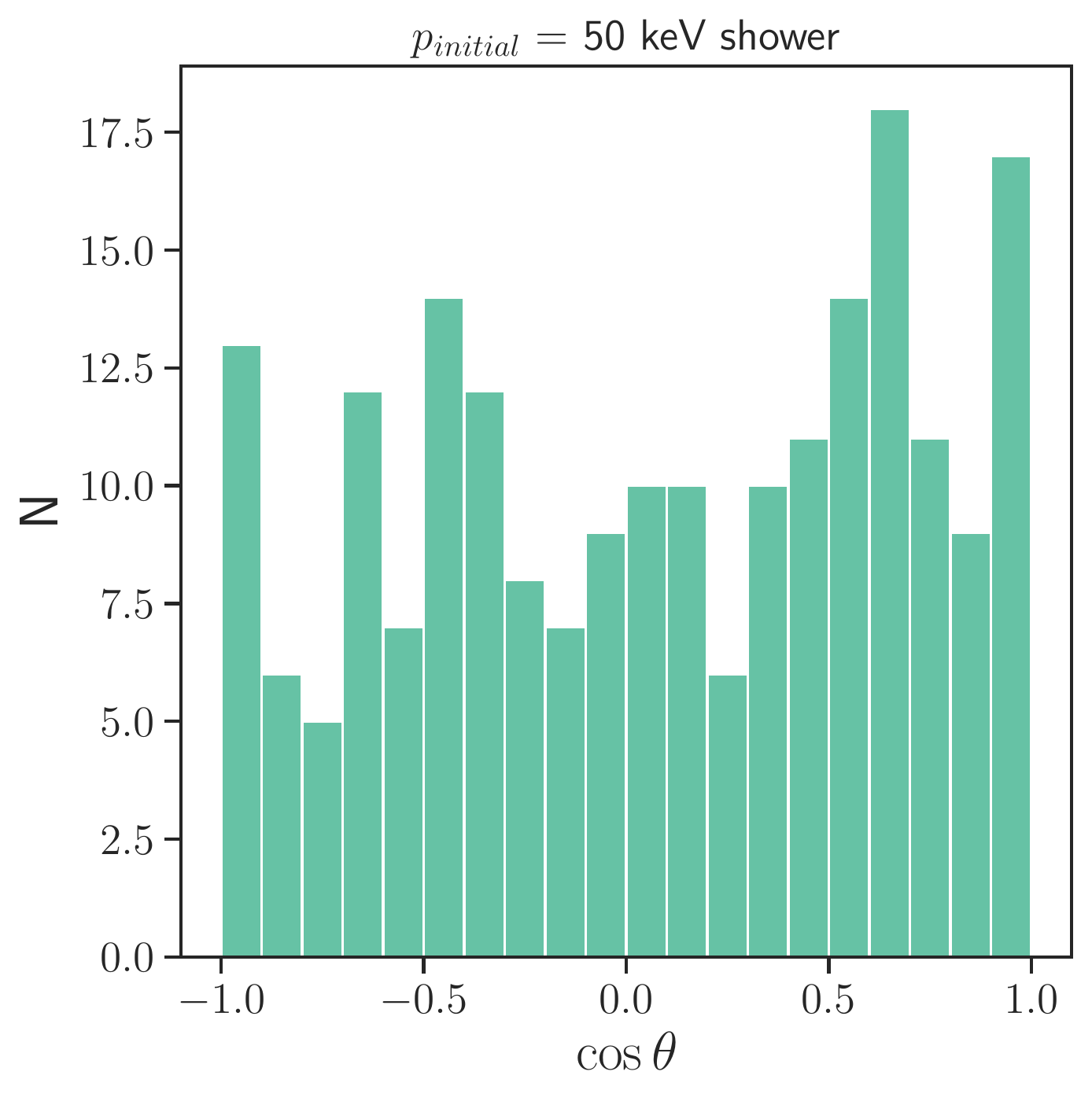}~~~~~~
\includegraphics[width=0.31\columnwidth]{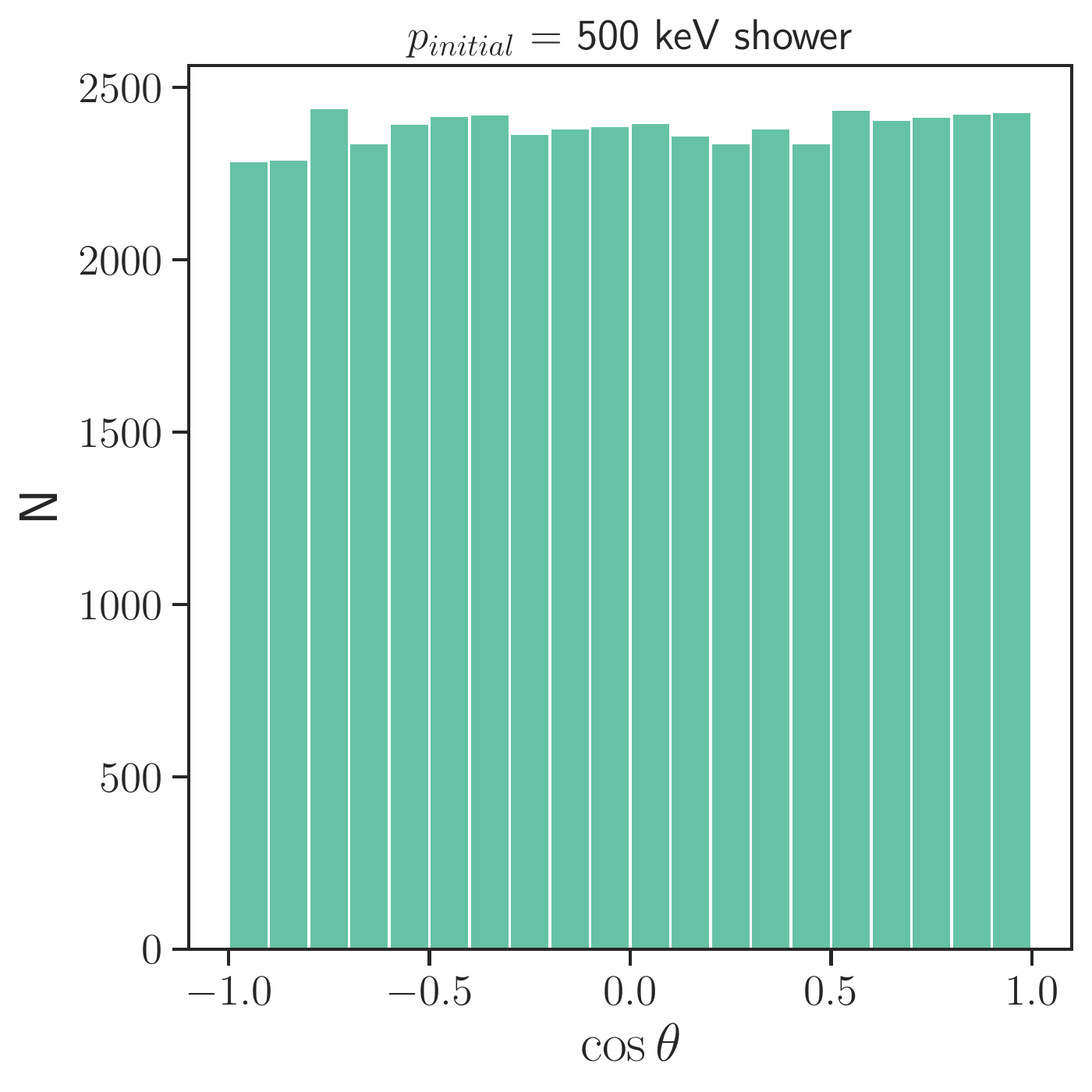}
\caption{Typical quasi-particle spectra from our simulation. In the first (second, third, fourth) row of panels we show distributions of quasi-particle momenta (energies, velocities, $\cos\theta$), for an initial He atom momentum of 50 keV (left column) and 500 keV (right column).}
\label{fig:spectra}
\end{figure}

In \cref{fig:spectra} we show typical quasi-particle spectra from our simulation. In the first (second, third, fourth) row of panels we show distributions of quasi-particle momenta (energies, velocities, $\cos\theta$), for an initial He atom momentum of 50 keV (left column) and 500 keV (right column). The top panels show that the number of phonons is significantly lower (compared to the number of rotons) due to the phase space suppression. Nonetheless, quasi-particle decays to softer modes, not included in our simulation, will eventually populate that region. The panels in the second row reveal that the energy spectrum starts with a peak at the gap energy $\Delta=0.75$ meV which corresponds to the local minimum of the roton dispersion. This is also reflected in the velocity graphs on the third row, where the peak around zero velocity is composed of slow rotons and maxons, as well as slow quasi-particle modes above $\sim 5$ keV. The plots in the last row demonstrate that the resulting distribution is almost isotropic. The discussion in \cref{sec:IntofQPs} below will show that all these quasi-particles will become thermalized. Therefore, instead of using the generated quasi-particles from \cref{fig:spectra}, in the later sections we shall sample the quasi-particle  spectrum from a Bose-Einstein distribution.  

\section{Effects of interactions among quasi-particles}

In this section we investigate the effects of interactions between the quasi-particles produced at the end of the cascade. These interactions may thermalize the ensemble of quasi-particles. In order to determine whether this actually occurs, we first do a back-of-the-iphone estimate comparing the quasi-particle interaction rate $n\sigma v$ and the quasi-particle production rate $\Gamma_\text{inel}$, for a given recoil energy or momentum. After that, we present the theoretical solution of last scattering surface and the thermal temperature. More accurate results from our simulation are presented in \cref{sec:simulationT}.

\subsection{Theoretical overview} 
\label{sec:IntofQPs}

We consider the two well known types of excitations with analytic dispersion relations -- phonons and rotons. The ``phonon" refers to a quasi-particle with momentum below $\sim 1.2$ keV; its dispersion is linear with a small cubic correction:
\begin{equation}
E_{\text{phonon}}\simeq c_{s}(p-\frac{\gamma}{\Lambda^{2}}p^{3}),\label{eq:PhononDis}
\end{equation}
where $c_{s}=240\,\text{m}/\text{s}$ is the sound speed, $\Lambda=(\rho c_{s})^{1/4}$
is the UV scale of the superfluid, and $\gamma\sim {\cal O}(1)$ is a dimensionless
parameter such that $\gamma/\Lambda^{2}=0.27\textup{~\AA}^2$. The ``roton" is a quasi-particle of momentum $\sim p_\ast=3.84$ keV at the local minimum of the dispersion curve. Its energy is parameterized as
\begin{equation}
E_{\text{roton}}\simeq\Delta+\frac{(p-p_{*})^{2}}{2m_{*}},\label{eq:RotonDis}
\end{equation}
where $m_{\ast}\simeq0.16m_{\text{He}}$
is the effective roton mass.

Several relevant interactions are well studied in the literature \cite{landau1941theory,landau1949theory,landau1987statistical,Nicolis_2018,1965511,PhysRevLett.119.260402,Matchev:2021fuw}, including phonon decay, phonon 2-2 scattering, roton 2-2 scattering, and phonon-roton 2-2 scattering\footnote{Phonon-roton scattering is the most complicated among these processes. There is existing controversy over the leading orders of cross section in the scenario that the initial roton is not at the exact bottom of the dispersion curve. In a following theory project, we will elaborate on the novel development of the last process, and discuss the different results involving initial phonon and roton states.}. In Table~1 of \cite{Matchev:2021fuw}, we listed the main results for these interaction cross sections. Using the parameter values of eqns.~(\ref{eq:PhononDis}) and (\ref{eq:RotonDis}), we find that the cross sections are of similar magnitude,  $\sigma\,v\sim 10^{-6} - 10^{-7}\ \text{keV}^{-2}$. 

With those ingredients, we are now ready to check for thermalization. If at the end of the atomic cascade, when 
all quasi-particles have just been produced, the quasi-particles have already experienced multiple interactions, we can safely claim that the quasi-particle system is thermalized. We perform a back-of-the-envelope estimation as follows. 

From \cref{fig:DM-Helium recoil}, we know that the initial recoil momentum $P_\text{ini}$ of the helium atom triggering the cascade ranges from 1 to $10^3$ keV. According to \cref{fig:HeRad}, when the helium atom momentum drops to below $\sim 20$ keV, 
the atom will predominantly start to radiate quasi-particles, thus not affecting the number of atoms in the cascade. Therefore, energy conservation implies that a recoil momentum of $P_\text{ini}$ keV will  dissipate to $\left(\frac{P_\text{ini}}{20\  \text{keV}}\right)^2$ slower helium atoms. Each of those slow atoms in turn will radiate about 100 quasi-particles, assuming that the quasi-particles' energies are $\sim 1$ meV. We take the typical scattering and radiation rate from \cref{fig:HeRad} as $10^{13}$ s$^{-1}$. The radiation of all quasi-particles will take $\Delta t \sim 100\times 10^{-13}\ \text{s}=10^{-11}$ s, which is longer than the prior atomic cascade time because the number of atoms increase exponentially with time during a cascade. Therefore, we estimate the total time of cascade and radiation to be $10^{-11}$ s.

During this time, the quasi-particles (with velocity ${\cal O}(100)\, \text{m}/\text{s}$) may expand up to $\Delta R \sim 100\  \text{m}/\text{s} \times 10^{-11}\  \text{s} = 10^{-9} \ \text{m} = 1$ nm. We then estimate the the interaction rate $n\sigma v$ for quasi-particle self-interactions as
$$
n\sigma v \sim\frac{\left(\frac{P_\text{ini}}{20\  \text{keV}}\right)^2 \times 100\times \sigma\,v}{\frac{4\pi}{3} (\Delta R)^3} \sim
\left(\frac{P_\text{ini}}{20\  \text{keV}}\right)^2 \times 10^{11}\ \text{s}^{-1}.
$$
We see that the corresponding timescale ranges from $10^{-14}$ s (for $P_\text{ini}\sim 1$ MeV) to $10^{-11}$ s (for $P_\text{ini}\sim 20$ keV). This timescale is smaller than the timescale of $10^{-11}$ s for producing the quasi-particle shower. Therefore, quasi-particles dissipated from $P_\text{ini}\gg 20$ keV recoil momentum will interact with each other multiple times, i.e. become fully thermalized, by the time when all quasi-particles are produced. When considering recoil momenta $\lesssim 20 \, {\rm keV}$, only a few quasi-particles are produced at this scale, and they infrequently interact with each other. Thus we treat these quasi-particles as free-streaming from the beginning.

\subsection{Quasi-particle thermalization}
\label{sec:Thermalization}

In the following, we present our procedure to derive the thermalized distribution of quasi-particles. First of all, we estimate the last scattering surface of the final quasi-particle \enquote{plasma} system, assuming an isotropic configuration distribution. Borrowing from the analogous concept in Cosmology \cite{Dodelson:2003ft,Weinberg:2008zzc}, at the last scattering surface, the quasi-particle interaction rate equals the expansion rate. Modeling the isotropic quasi-particle \enquote{plasma} as a sphere, the last scattering radius $R_{\text{ls}}$ is determined by the radius for which the optical depth is unity: 
\begin{equation}
   \tau=\int_{R_{\text{ls}}}^{\infty}\frac{dr}{(n\sigma)^{-1}}=1,\quad n=\frac{N}{V},
   \label{eq:expansion}
\end{equation}
where $N$ is the total number of quasi-particles, $\sigma$ is their interaction rate, and $V=4\pi r^3/3$.
 The solution of the previous equation gives the last scattering radius $R_\text{ls}=\sqrt{3N\sigma/8\pi}$. 
 $N$ is estimated by assuming that all final quasi-particle excitations have energy 1 meV: $N=\frac{P_\text{ini}^2/2m_\text{He}}{1\ \text{meV}}$.
 Plugging in the numerical values from the previous subsection, we find the magnitude of the last scattering radius $R_\text{ls}$ to be ${\cal O}(1)$ to ${\cal O}(10)$ nm, smaller than the distance between the sensors in realistic detectors. Therefore, the relevant distribution detected by the sensors is the thermalized spectrum of quasi-particles. The number of quasi-particles is not fixed, but can change due to a) decays of unstable phonons (with momenta below $1.2$ keV) to other phonons; b) decays of other quasi-particles to stable quasi-particles with momenta between $1.2$ and $4.6$ keV \cite{Hertel:2018aal,donnelly1981specific}; and c) multiple quasi-particle interactions which can change the number of quasi-particle modes. Therefore, the literature treats the chemical potential of quasi-particles as zero. Then we can express the Bose-Einstein distribution of the quasi-particles as
\begin{equation}
    n_\text{B-E}(\vec{p})= \frac{1}{\exp\frac{\omega(p)}{k_B T} - 1}.
\end{equation}
Using energy conservation, the initial recoil energy equals the total energy of the Bose-Einstein ensemble:
\begin{equation}
   \frac{P_\text{ini}^2}{2m_\text{He}}=\frac{4}{3}\pi R_\text{ls}^3 \int_0^{4.6\text{keV}}\frac{d^3p}{(2\pi)^3}\,\omega(p)\,n_\text{B-E}(p),
   \label{eq:solveT}
\end{equation}
where $\frac{4}{3}\pi R_\text{ls}^3$ is the volume of the thermal system. The momentum integration runs from 0 to 4.6 keV because we assume only phonons and stable quasi-particles exist in the thermalized distribution. 
\begin{figure}[tb]
\centering
\includegraphics[width=0.8\columnwidth]{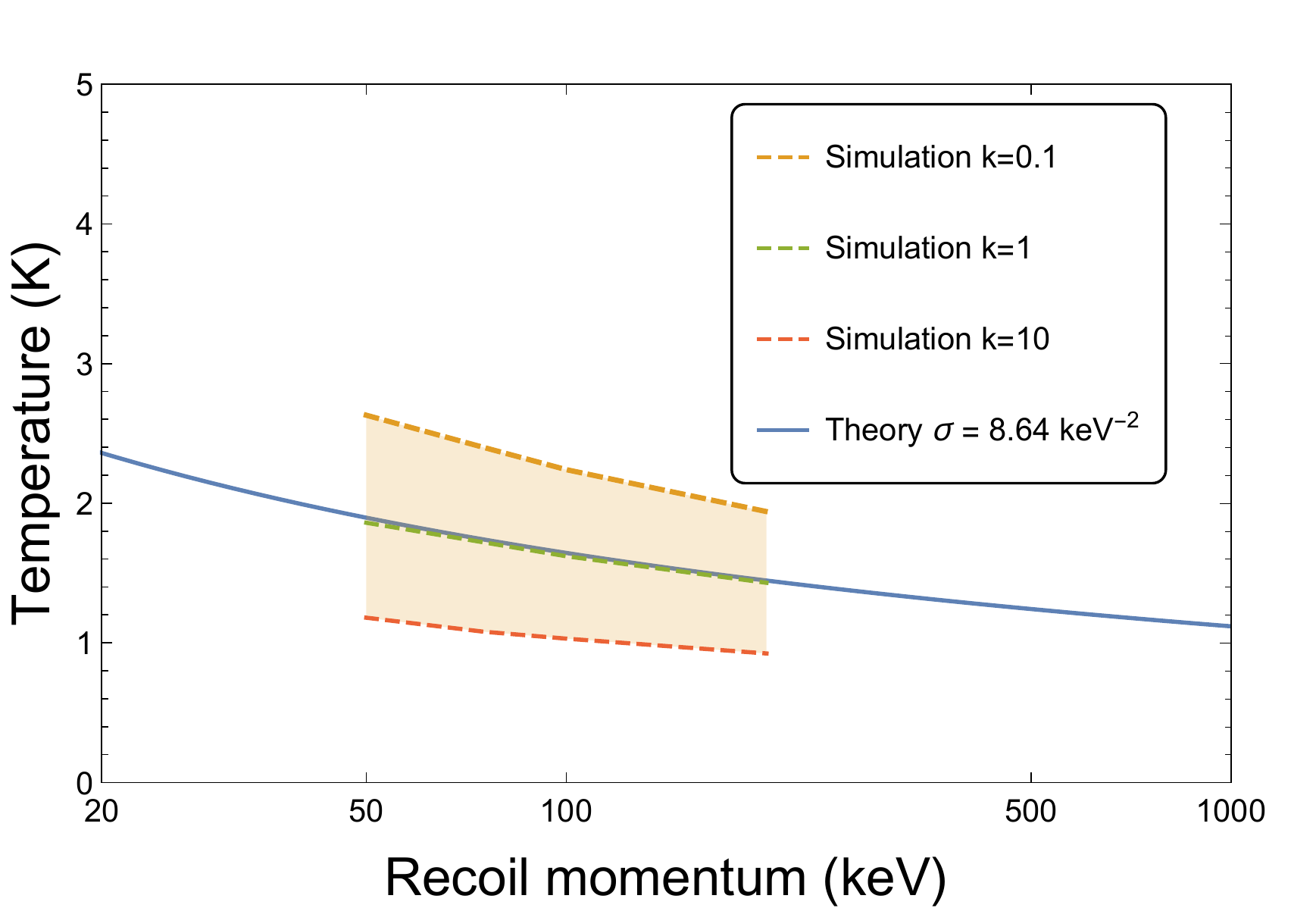}
\caption{The thermal temperature $T$ predicted from \cref{eq:solveT} as a function of the initial recoil momentum $P_\text{ini}$. The blue solid line corresponds to taking $\sigma=8.64\,\text{keV}^{-2}$ (the value of roton-roton average cross section provided by the simulation) when solving for the last scattering radius $R_\text{ls}$ appearing in \cref{eq:solveT}. The dashed lines in the range 50 to 200 keV are obtained by using the simulated average $\sigma$, with the phonon-roton scattering cross section multiplied by a $k$-factor of  $k=1$ (green), $k=0.1$ (orange), or $k=10$ (red).}
\label{fig:Thermal T}
\end{figure}

\subsection{Monte Carlo simulation and the thermal temperature}
\label{sec:simulationT}

We can use our MC simulations described in \cref{sec:MCsimulation} to provide a numerical estimate for $T$ from the previous subsection. The simulation provides averaged values of the different scattering cross sections between quasi-particles of all types. These cross-sections can be used in \cref{eq:expansion} for the calculation of $R_\text{ls}$, which can then be substituted in \cref{eq:solveT} to solve for $T$. The result is shown as the blue solid line in \cref{fig:Thermal T}, for which we have used the value $\sigma=8.64\ \text{keV}^{-2}$ suggested by our simulations. As mentioned in \cref{sec:IntofQPs}, the region of $P_\text{ini}$ below $20$ keV is unnecessary for our simulation, since the recoiled helium atom will radiate quasi-particles without an atomic cascade, and the number of quasi-particles will be insufficient for thermalization.

 One potential problem with our simulations is that the expression for the phonon-roton scattering cross section is only valid for phonons that have a much smaller momentum than the roton \cite{Matchev:2021fuw,Nicolis_2018,PhysRevLett.119.260402}. Therefore, we cannot reliably sample the cross section between hard phonons and rotons.
In order to get an idea of the effect from this theoretical uncertainty, we introduce a $k$-factor for the phonon-roton cross section and in \cref{fig:Thermal T} show results for  $k=10$ (red dashed line), $k=1$ (green dashed line) and $k=0.1$ (orange dashed line). The width of the shaded band enclosed between the orange and red dashed lines is indicative of the corresponding uncertainty on the derived thermal temperature $T$, which should be kept in mind when discussing projections for the experimental sensitivity below.

\section{Experimental signals} 
\label{sec:flux&force}

\begin{figure}[tb]
\centering
\includegraphics[width=0.7\columnwidth]{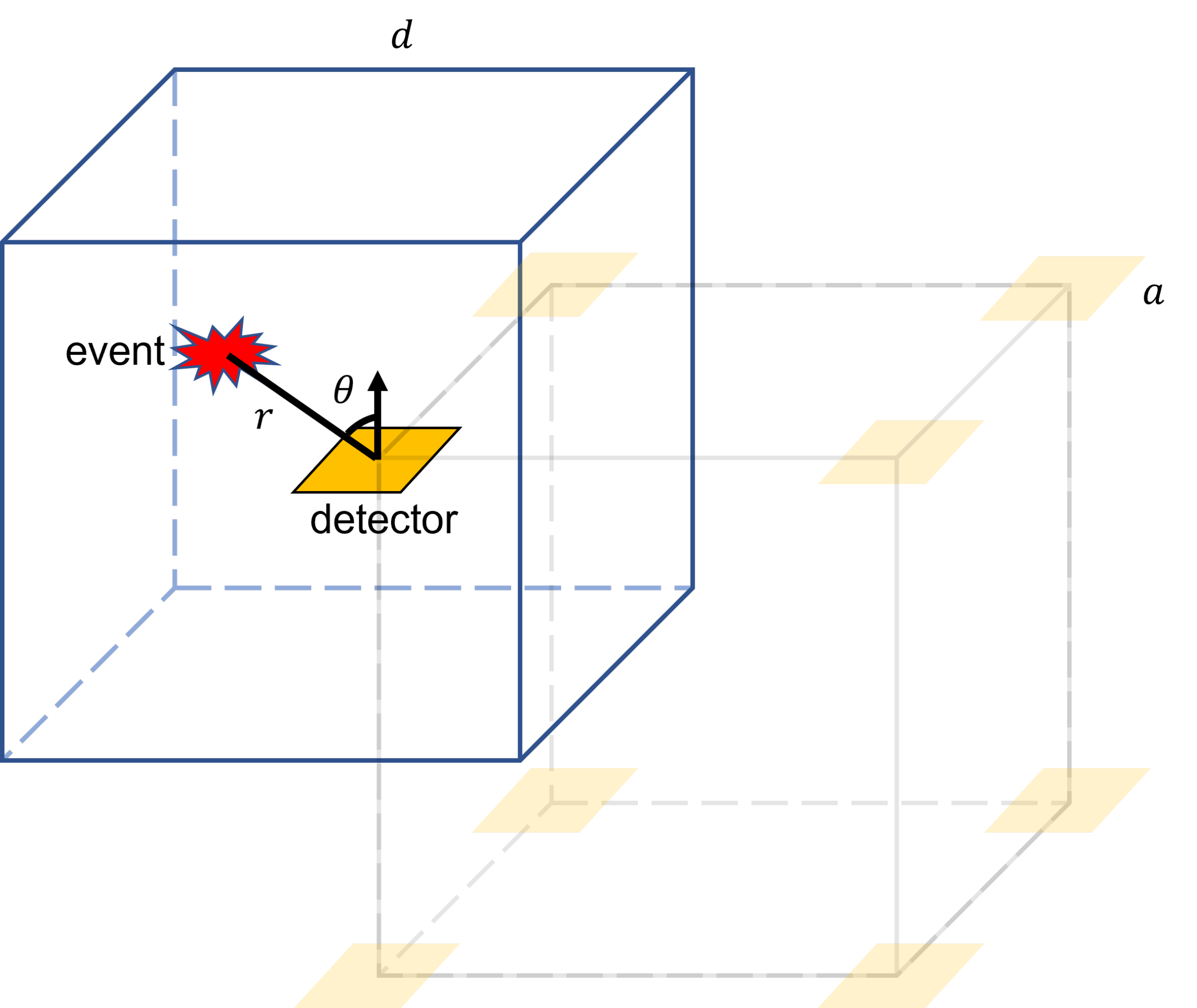}
\caption{Illustration of the geometry used to calculate the signal on the detector surface. We assume that a single sensor is a square of size $a \times a$, and the sensors are arranged in a 3D lattice with spacing $d$. The event can happen at any location in the lattice, but it will be first detected by the closest sensor. For one given sensor, we thus calculate the signal from events within a cube of volume $d^3$. See \cref{sec:flux&force} for details. 
}
\label{fig:IllustrationRad}
\end{figure}

The previous analysis and simulations are applicable to various types of dark matter searches in superfluid helium.
One such proposal is the HeRALD experiment \cite{Hertel:2018aal}, which tries to detect dark matter scattering events occurring in the bulk of the superfluid by sensing quantum evaporation of helium atoms from the superfluid surface.
Here we study the detection prospects of mechanical oscillators like NEMS, which instead sense the force (or momentum deposit) due to quasi-particles. 
For a given momentum threshold and detector size, we deduce the potential to discover the dark matter signal using mechanical oscillators and
compare with the HeRALD experiment, leaving the detailed experimental design and study of the backgrounds for a future work \cite{MSXY}.

\paragraph{Detector setup.} In order to study the quasi-particle flux onto a generic device, we consider a simple detector model as follows. We assume a large supply of 2D square-shaped sensors of area $a \times a$, which are  placed on a 3D grid inside a superfluid target, with an inter-sensor distance of $d$. The distance between the event and the closest target sensor is up to half of a diagonal $\sqrt{3}d/2$. Each sensor at the cell vertex first receives the signal from its own octic cube. Therefore, we simplify the geometry and study the events happening within a cube of volume $d^3$ with a sensor at the center (see \cref{fig:IllustrationRad}). 

\begin{figure}[tb]
\centering
\includegraphics[width=0.8\columnwidth]{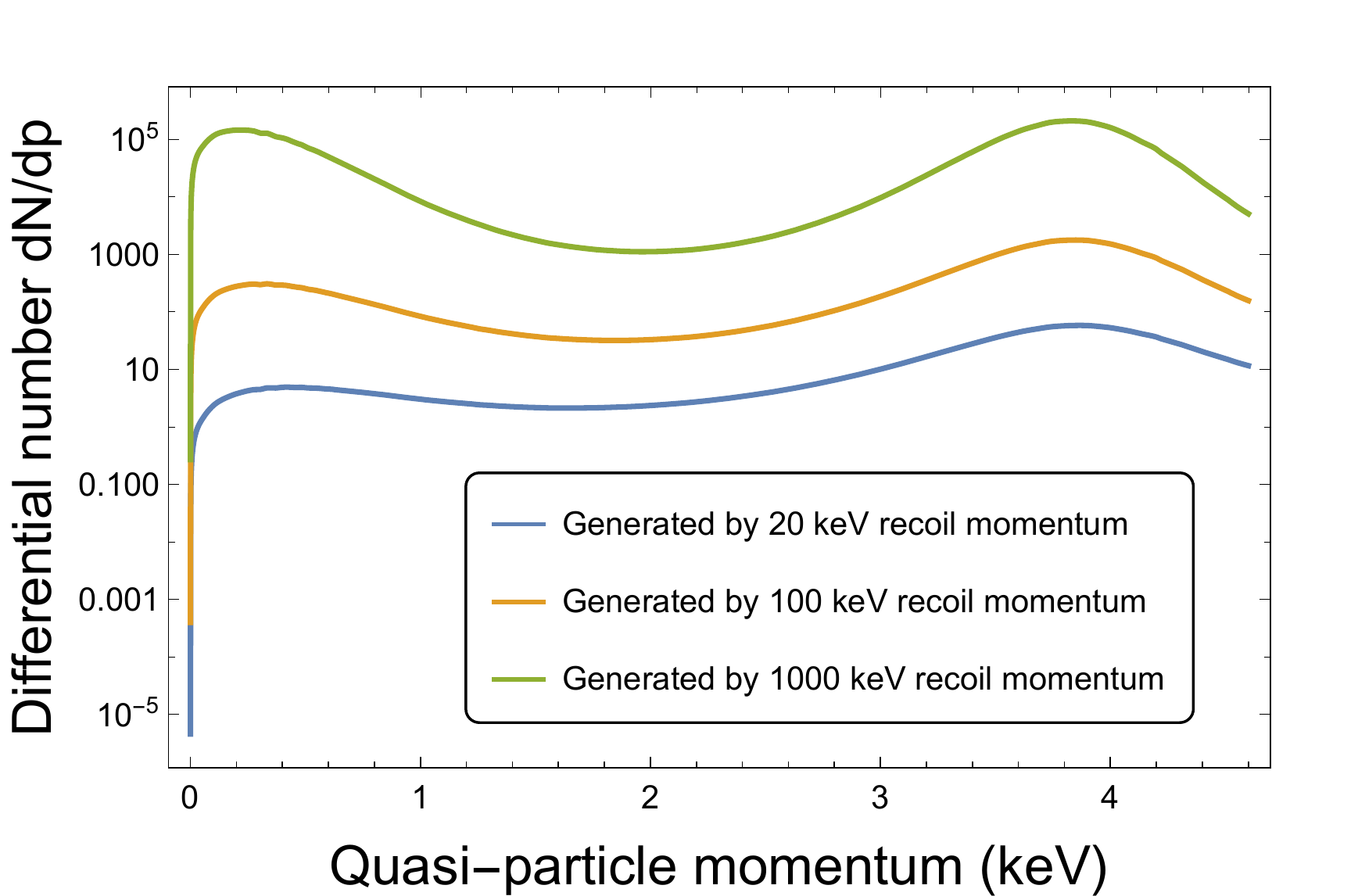}
\caption{The spectrum of differential particle number $dN/dp$ as a function of quasi-particle momentum $p$. The thermalized quasi-particle systems are generated by a recoil momentum of 20 keV (blue), 100 keV (orange), and 1000 keV (green) respectively. The total number of quasi-particles is the area under the curves. We assume only phonons and stable quasi-particles  exist in the thermalized distribution.}
\label{fig:Thermal Quasi-particle number profile}
\end{figure}

We also model the detection as an  event free streaming to the sensor of area $a^2$ from a distance $r$. Assuming the equilibrium quasi-particle ensemble to be isotropic (which is confirmed in our simulations), the probability of a single given particle reaching the surface is given by the ratio between its solid angle to the detector $\Omega(\theta,\phi,r,a)$ and the total solid angle $4\pi$. The solid angle $\Omega(\theta,\phi,r,a)$ belongs to a leaning pyramid configuration and is solved numerically using Mathematica build-in commands. If $P_{\rm ini} > 20 \, {\rm keV}$, the quasi-particles reach thermalization. Then we calculate the differential profile, $dN/dp$, 
the number of quasi-particles with given momentum $p$, according to a thermal distribution:
\begin{equation}
    \frac{dN}{dp}=\frac{V}{2\pi^2}\frac{p^2}{\exp{\frac{\omega(p)}{T}}-1},
    \label{eq:dn/dp}
\end{equation}
where $V$ is the last scattering volume estimated from \cref{eq:expansion}, while $T$ is the thermal temperature discussed in \cref{fig:Thermal T}. Results for a few representative values of $P_{\text ini}$ are shown in \cref{fig:Thermal Quasi-particle number profile}.
For $P_{\rm ini} < 20 \, {\rm keV}$, the quasi-particles are free-streaming from their point of origin. 
We estimate the number of events using energy conservation. According to the simulation results in \cref{sec:MCsimulation}, for the estimation of the quasi-particle flux we further assume that most of the quasi-particles are rotons.

\begin{figure}[t]
  \centering
  \subfloat[Probability~(\ref{eq:PofO1keV}) of 1 keV threshold]{\includegraphics[width=0.5\textwidth]{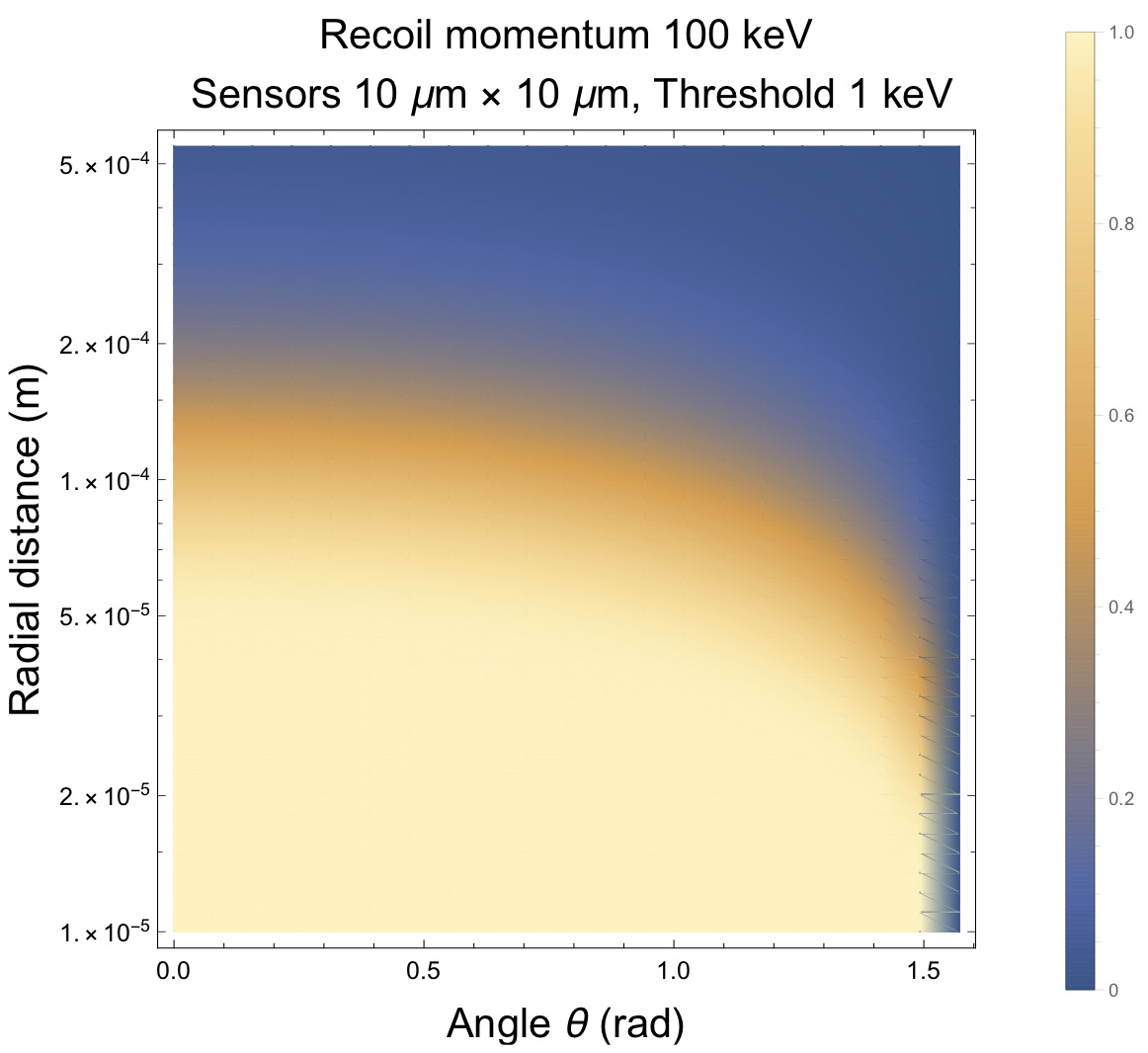}\label{fig:pgraph1}}
  \hfill
  \subfloat[Probability~(\ref{eq:PofO10keV}) of 10 keV threshold]{\includegraphics[width=0.5\textwidth]{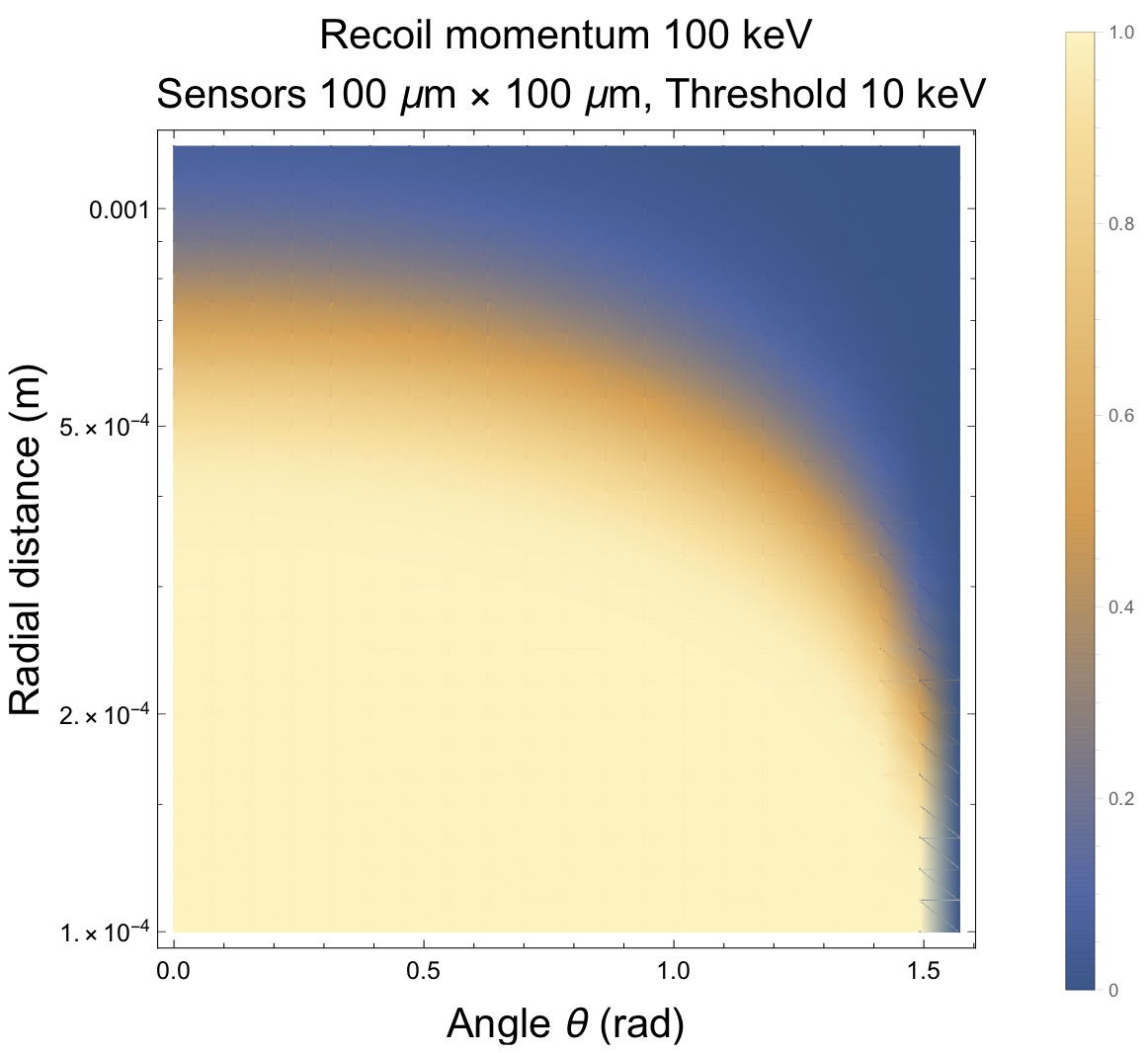}\label{fig:pgraph2}}
  \caption{Detection probability of two example sensors. (a) The probability of a 1 keV threshold sensor from \cref{eq:PofO1keV}. The initial recoil momentum from the DM is 100 keV, the sensor area is $10\times 10\,\mu \text{m}^{2}$. (b) The probability of a 10 keV threshold sensor from \cref{eq:PofO10keV}. The initial recoil momentum from the DM is 100 keV, the sensor area is $100\times 100\,\mu \text{m}^{2}$. Here we vary the zenith angle $\theta$ along the horizontal axis, while setting the azimuth angle $\phi=0$. The probability increases as the event radial distance decreases and as the direction gets closer to the normal ($\theta=0$).}
  \label{fig:probabilitygraph}
\end{figure}

\paragraph{${\cal O}(1)$ keV threshold.} Sensors of momentum threshold $\sim {\cal O}(1)$ keV are capable of detecting single quasi-particles. Taking 1 keV threshold as an example, as long as at least one quasi-particle reaches the sensor, the event can be counted as detectable. There is a higher probability of detection when the particles are streaming from a closer distance and a normal angle from the sensor. We thus extract this probability between the event and the nearest sensor:
\begin{equation}
    P(\theta,\phi,a,r)\,=\,1-\left(1-\frac{\Omega(\theta,\phi,a,r)}{4\pi}\right)^{N(p_\text{min})}.
    \label{eq:PofO1keV}
\end{equation}
    Here $p_\text{min}$ is the threshold momentum of the sensor, and $N(p_\text{min})$ is the total number of quasi-particles with momentum above threshold, which is integrated from \cref{eq:dn/dp}. We plot an example density plot for this probability function in the left panel of \cref{fig:probabilitygraph}. The number of quasi-particles arriving at the sensor decreases when the events happen further from the sensor and away from its normal direction. 

\begin{figure}[t]
  \centering
  \subfloat[Spatial efficiency of 1 keV threshold]{\includegraphics[width=0.5\textwidth]{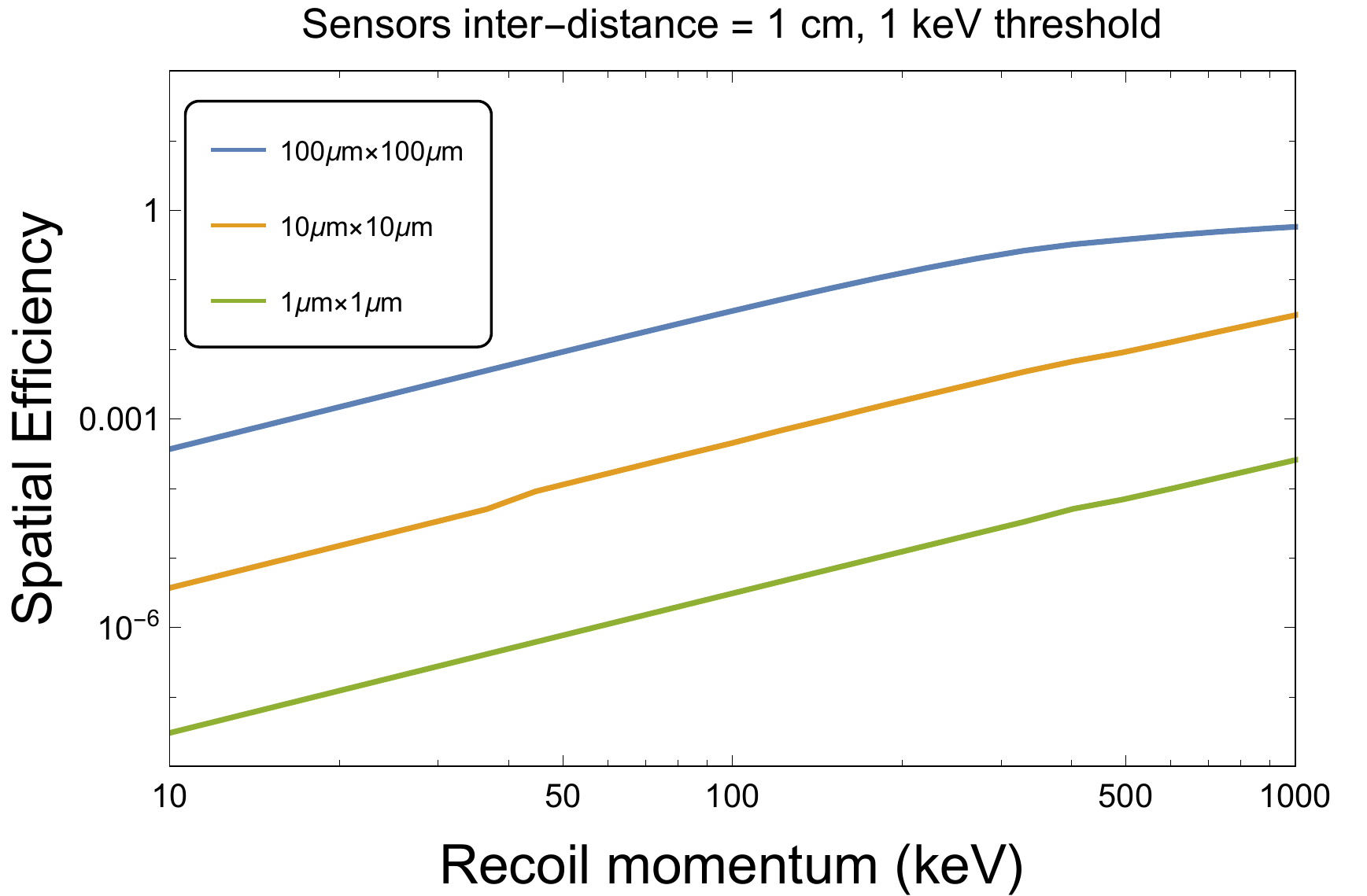}\label{fig:SpatialEff1}}\hfill \subfloat[Spatial efficiency of 10 keV threshold]{\includegraphics[width=0.5\textwidth]{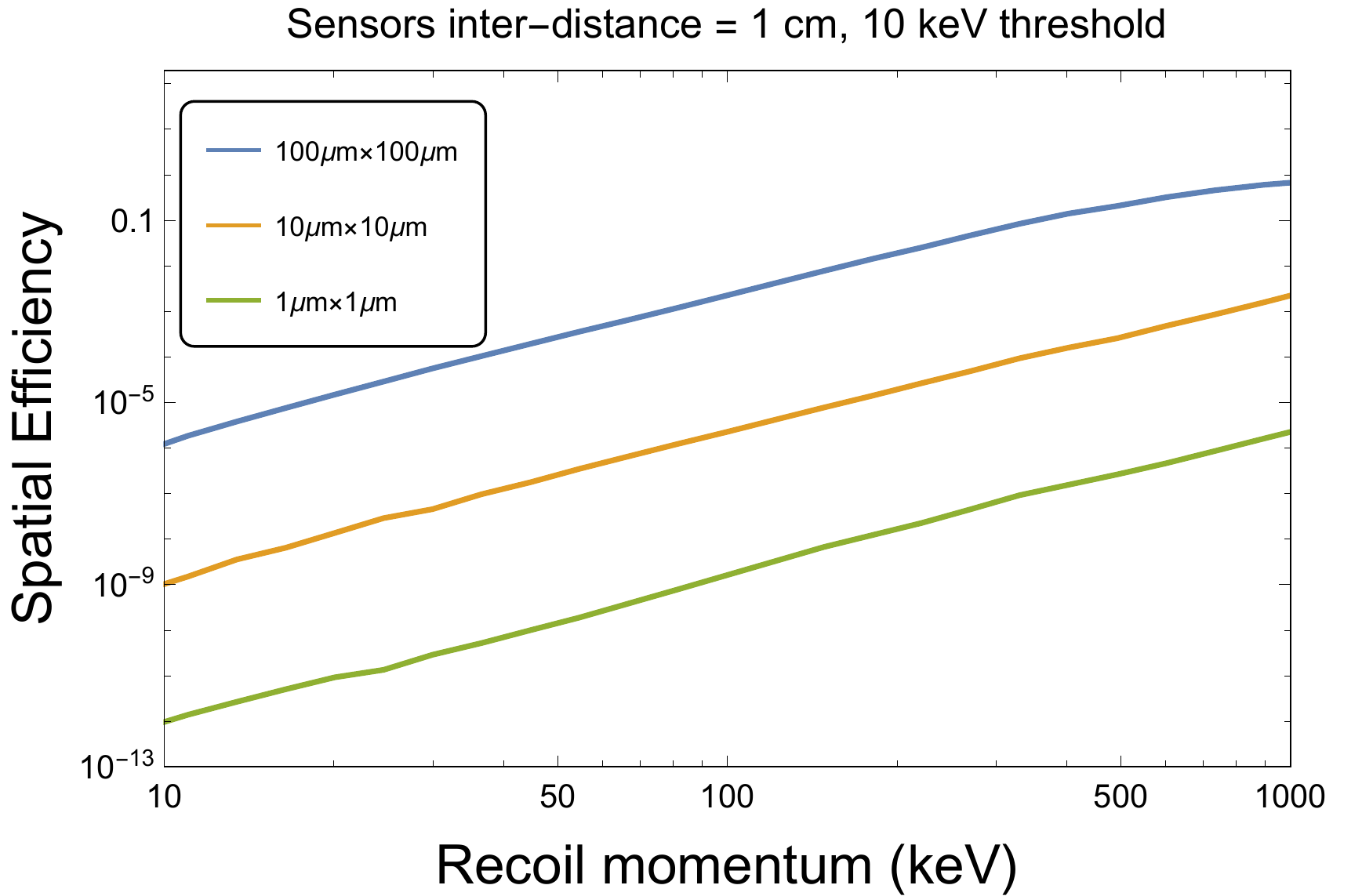}\label{fig:SpatialEff10}}
  \caption{Spatial efficiency (\ref{eq:EFofO1keV}) of various detector setups. (a) Inter-distance between the sensors is 1 cm and the momentum threshold is 1 keV. The blue (orange, green) curve represents the spatial efficiency of a sensor with area of $100 \times 100\,\mu m^2$ ($10 \times 10\,\mu m^2$, $1 \times 1\,\mu m^2$). (b) The spatial efficiency when the momentum threshold is 10 keV.}
   \label{fig:DepEff}
\end{figure}

Since the DM-Helium scattering rate eq.~(\ref{eq:differentialrate}) is independent to the coordinate space, we average the probability $P(\theta,\phi,a,r)$ over the whole sensor cell and call it the Spatial Efficiency (SE):
\begin{equation}
    {\cal E}_S\,=\,\frac{1}{d^3}\int^{d/2}_{-d/2}dx\int^{d/2}_{-d/2}dy\int^{d/2}_{-d/2}dz\,P(\theta,\phi,a,r),
    \label{eq:EFofO1keV}
\end{equation}
where radial distance $r=\sqrt{x^2+y^2+z^2}$, the zenith angle $\theta=\tan^{-1}(\sqrt{x^2+y^2}/z)$, and the azimuth angle $\phi=\tan^{-1}(y/x)$. The spatial efficiency eventually depends on the recoil momentum $P_\text{ini}$, the sensor distance $d$, and the sensor area $a^2$. We show the spatial efficiency function \cref{eq:EFofO1keV} results for ${\cal O}(1)$ keV threshold scenario in \cref{fig:SpatialEff1}. Comparing the three curves, we notice that the spatial efficiency is generally proportional to the area $a^2$ of the sensor. This is because the probability (\ref{eq:PofO1keV}) takes the limit $N(p_{\min})\Omega(\theta,\phi,a,r)/4\pi$ at a large particle number $N(p_{\min})\gg 1$. 
In addition to the nearest sensor, the other sensors further away will also receive some signal, which we roughly estimate to be on the order of $40\%$ (for a threshold of 1 keV), but to be conservative, this correction will not be included in our sensitivity projections below.

\paragraph{${\cal O}(10)$ keV threshold and beyond.} Sensors of momentum threshold $\sim {\cal O}(10)$ keV are incapable of detecting a single quasi-particle. Taking 10 keV threshold as an example, we estimate that it takes $\ge 3$ quasi-particles reaching the sensor to trigger a detectable event. The probability (\ref{eq:PofO1keV}) can thus be generalized into a cumulative Binomial distribution:
\begin{equation}
    P(\theta,\phi,a,r)\,=\,1-\sum_{m=0}^{2}C_{N}^{m}\left(1-\frac{\Omega(\theta,\phi,a,r)}{4\pi}\right)^{N-m}\left(\frac{\Omega(\theta,\phi,a,r)}{4\pi}\right)^m.
    \label{eq:PofO10keV}
\end{equation}
Here the number $N$ is the total number of quasi-particles without the cap of $p_\text{min}$. Similar formalism can be applied to ${\cal O}(100)$ keV threshold. For example, we estimate that at least 25 quasi-particles are required to trigger a 100 keV threshold sensor. Therefore, the cumulative sum in \cref{eq:PofO10keV} is up to $m=24$. The total number of quasi-particles $N$ must exceed the minimum quasi-particle number that triggers the threshold, which is the maximum value of $m$ in this equation. For example, there are $\lesssim 3$ quasi-particles produced below 5 keV recoil momentum (no matter whether thermalized or not), thus the probability and the spatial efficiency vanishes.

Despite these changes, the spatial efficiency function follows the same formalism (\ref{eq:EFofO1keV}). As shown in \cref{fig:SpatialEff10}, the spatial efficiency in this scenario is about three orders of magnitude smaller than the previous case. Moreover, the spatial efficiency no longer scales with the area but with $a^3$. It is therefore unnecessary to compute the contribution from the sensors further away because they are unable to compensate for the loss of sensitivity.

\paragraph{Sensitivity projection.} DM with a given mass could generate a recoil momentum of any value with the rate distribution shown in \cref{fig:DM-Helium recoil}. The momentum threshold and other parameters of the detector determine the probability of an event at location  $(\theta, \phi, r)$ being detectable. Mathematically, the integration of $d\Gamma/dq^2$ from \cref{fig:DM-Helium recoil} is thus weighted by the curves in the \cref{fig:DepEff} when we calculate the total event rate. Although the spatial efficiency is a continuous function, the total detectable event rate vanishes when the total number of quasi-particles is less than the minimum requirement to trigger the threshold, as mentioned above.

\begin{figure}[tb]
\centering
\includegraphics[width=0.8\columnwidth]{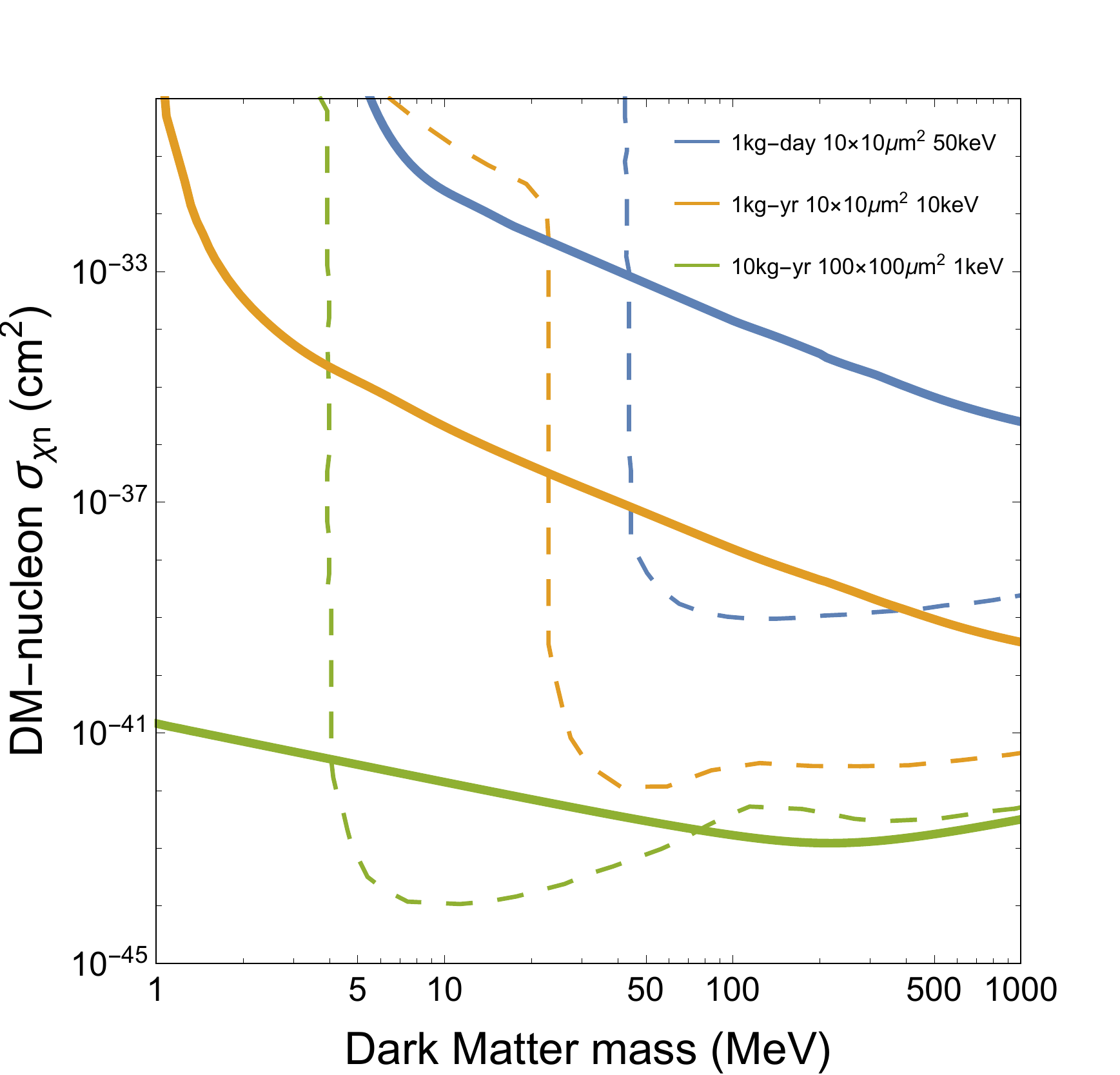}
\caption{The solid lines show 95\% C.L. exclusion regions for our proposed experiment for different detector parameters (assuming zero background). The legends label the three detector parameters in the following order: the exposure time and target mass, the area of the sensor, and the momentum threshold. The distance between two sensors is set as 1 cm. We also include the HeRALD results \cite{Hertel:2018aal} (dashed lines) which consider the solar neutrino background. The parameters for the HeRALD results are: 1 kg-day exposure with 40 eV energy threshold (blue), 1 kg-yr exposure with 10 eV energy threshold (orange), and 10 kg-yr exposure with 0.1 eV energy threshold (green).
}
\label{fig:Money plot}
\end{figure}

In \cref{fig:Money plot}, we plot the estimated reach in DM-Nucleus recoil cross section $\sigma_{\chi n}$ for several detector configurations, based on 90 percent confidence level (2.3 events) per exposure time per target mass \cite{Schutz:2016tid,Feldman:1997qc,Lista:2016chp,Bhat:2010zqs,Sinervo:2002sa,Barlow:2003cx}, assuming zero background. For a one year exposure with one kilogram of target, the constraint on the DM-Helium scattering rate is:
\begin{equation}
    \frac{1}{M_\text{target}}\times\int_{q_{\min}}^{q_{\max}}   {\cal E}_S (q) \, \frac{d {\rm R}}{d\textbf{q}^2}2q\,dq  \, <0.728\times 10^{-7}\text{s}^{-1}\text{kg}^{-1},
\end{equation} 
where $q_{\min}$ is the cutoff by $N>1$ in \cref{eq:PofO1keV} or $N>m_{\max}$ in \cref{eq:PofO10keV}, and $q_{\max}$ is the recoil momentum upper limit from classical kinematics. The variation of exposure includes per day per kilogram, per year per kilogram, and per year per 10 kilogram, with the expected number of events scaling accordingly. A combination of smaller distance between sensors, larger area of the sensor, and longer exposure time allows to probe smaller cross sections.


\section{\bf Conclusions} 
\label{sec:conclude}

We have proposed a new method of detecting sub-GeV DM using the spectrum of the quasi-particle excitations in superfluid $^4$He generated as a result of a DM collision with a He nucleus. The key idea is to leverage modern force-sensitive devices, such as NEMS oscillators, to detect the momentum flux of the quasi-particles.
With a kg-year exposure, we have demonstrated that this superfluid experiment can strongly constrain the DM-Nucleus interaction within the sub-GeV mass range. Our study is also the first to theoretically study (1) the production of quasi-particles from the resulting helium atom cascade; (2) the thermalization and decoupling of quasi-particles; and (3) the temperature of the thermalized quasi-particle system in a superfluid DM detector. The findings include:
\begin{itemize}
    \item 
The DM collision initiates a cascade of helium atoms which gradually lose momentum by elastic scattering, and later, once they reach ${\cal O}(10)$ keV, by radiation of quasi-particles. 
    \item Quasi-particles generated by a recoil momentum $P_\text{ini}\gtrsim {\cal O}(10)$ keV will thermalize as a result of their self-interactions. For $P_\text{ini}\lesssim {\cal O}(10)$ keV, the quasi-particles are free streaming from the beginning. 
    \item The temperature of the thermalized quasi-particles system generated by sub-GeV DM is ${\cal O}(1)$ Kelvin. 
\end{itemize}
These results lead to a prediction of the detectability of events in the vicinity of a NEMS sensor. With a 10 kg-yr exposure and a 1 keV momentum threshold, the ideal detector setup is able to push the DM-nucleon exclusion region down to about $10^{-41}$ to $10^{-43}\,\text{cm}^2$.

\acknowledgments

We are grateful to Rasul Gazizulin, Wei Guo, Tongyan Lin and Bin Xu for useful discussions. This work was supported in part by the United States Department of Energy under Grant No. DE-SC0022148, and the National Science Foundation under Grant No. PHY-2110766.

\vspace{5mm}


\bibliographystyle{JHEP}

\bibliography{ref}

\providecommand{\href}[2]{#2}\begingroup\raggedright\begin{thebibliography}{100}

\bibitem{Zwicky:1933gu}
F.~Zwicky, \emph{{Die Rotverschiebung von extragalaktischen Nebeln}},
  \href{https://doi.org/10.1007/s10714-008-0707-4}{\emph{Helv. Phys. Acta}
  {\bfseries 6} (1933) 110}.

\bibitem{Bertone:2016nfn}
G.~Bertone and D.~Hooper, \emph{{History of dark matter}},
  \href{https://doi.org/10.1103/RevModPhys.90.045002}{\emph{Rev. Mod. Phys.}
  {\bfseries 90} (2018) 045002}
  [\href{https://arxiv.org/abs/1605.04909}{{\ttfamily 1605.04909}}].

\bibitem{Arbey:2021gdg}
A.~Arbey and F.~Mahmoudi, \emph{{Dark matter and the early Universe: a
  review}}, \href{https://doi.org/10.1016/j.ppnp.2021.103865}{\emph{Prog. Part.
  Nucl. Phys.} {\bfseries 119} (2021) 103865}
  [\href{https://arxiv.org/abs/2104.11488}{{\ttfamily 2104.11488}}].

\bibitem{Knapen:2017xzo}
S.~Knapen, T.~Lin and K.~M. Zurek, \emph{{Light Dark Matter: Models and
  Constraints}}, \href{https://doi.org/10.1103/PhysRevD.96.115021}{\emph{Phys.
  Rev. D} {\bfseries 96} (2017) 115021}
  [\href{https://arxiv.org/abs/1709.07882}{{\ttfamily 1709.07882}}].

\bibitem{lin2019tasi}
T.~Lin, \emph{{Dark matter models and direct detection}},
  \href{https://doi.org/10.22323/1.333.0009}{\emph{PoS} {\bfseries 333} (2019)
  009} [\href{https://arxiv.org/abs/1904.07915}{{\ttfamily 1904.07915}}].

\bibitem{Agnese_2019}
R.~Agnese, T.~Aralis, T.~Aramaki, I.~Arnquist, E.~Azadbakht, W.~Baker et~al.,
  \emph{Search for low-mass dark matter with cdmslite using a profile
  likelihood fit},
  \href{https://doi.org/10.1103/physrevd.99.062001}{\emph{Physical Review D}
  {\bfseries 99} (2019) }.

\bibitem{Armengaud_2019}
E.~Armengaud, C.~Augier, A.~Benoît, A.~Benoit, L.~Bergé, J.~Billard et~al.,
  \emph{Searching for low-mass dark matter particles with a massive ge
  bolometer operated above ground},
  \href{https://doi.org/10.1103/physrevd.99.082003}{\emph{Physical Review D}
  {\bfseries 99} (2019) }.

\bibitem{Aprile_2019}
E.~Aprile, J.~Aalbers, F.~Agostini, M.~Alfonsi, L.~Althueser, F.~Amaro et~al.,
  \emph{Light dark matter search with ionization signals in xenon1t},
  \href{https://doi.org/10.1103/physrevlett.123.251801}{\emph{Physical Review
  Letters} {\bfseries 123} (2019) }.

\bibitem{Aprile_2018}
{\scshape XENON Collaboration 7} collaboration, E.~Aprile, J.~Aalbers,
  F.~Agostini, M.~Alfonsi, L.~Althueser, F.~D. Amaro et~al., \emph{Dark matter
  search results from a one ton-year exposure of xenon1t},
  \href{https://doi.org/10.1103/PhysRevLett.121.111302}{\emph{Phys. Rev. Lett.}
  {\bfseries 121} (2018) 111302}.

\bibitem{Alkhatib_2021}
{\scshape SuperCDMS Collaboration} collaboration, I.~Alkhatib, D.~W.~P. Amaral,
  T.~Aralis, T.~Aramaki, I.~J. Arnquist, I.~Ataee~Langroudy et~al., \emph{Light
  dark matter search with a high-resolution athermal phonon detector operated
  above ground},
  \href{https://doi.org/10.1103/PhysRevLett.127.061801}{\emph{Phys. Rev. Lett.}
  {\bfseries 127} (2021) 061801}.

\bibitem{Agnes_2018}
{\scshape DarkSide Collaboration} collaboration, P.~Agnes, I.~F.~M.
  Albuquerque, T.~Alexander, A.~K. Alton, G.~R. Araujo, D.~M. Asner et~al.,
  \emph{Low-mass dark matter search with the darkside-50 experiment},
  \href{https://doi.org/10.1103/PhysRevLett.121.081307}{\emph{Phys. Rev. Lett.}
  {\bfseries 121} (2018) 081307}.

\bibitem{Abdelhameed_2019}
A.~Abdelhameed, G.~Angloher, P.~Bauer, A.~Bento, E.~Bertoldo, C.~Bucci et~al.,
  \emph{First results from the cresst-iii low-mass dark matter program},
  \href{https://doi.org/10.1103/physrevd.100.102002}{\emph{Physical Review D}
  {\bfseries 100} (2019) }.

\bibitem{Behr:2022tyz}
{\scshape ATLAS, CMS} collaboration, J.~K. Behr, \emph{{Searches for Dark
  Matter with the ATLAS and CMS Experiments using LHC Run 2
  (2015\textendash{}2018) Data}},
  \href{https://doi.org/10.22323/1.380.0137}{\emph{PoS} {\bfseries PANIC2021}
  (2022) 137}.

\bibitem{alexander2016dark}
J.~Alexander et~al., \emph{{Dark Sectors 2016 Workshop: Community Report}},  8,
  2016.

\bibitem{battaglieri2017us}
M.~Battaglieri et~al., \emph{{US Cosmic Visions: New Ideas in Dark Matter 2017:
  Community Report}},  in \emph{{U.S. Cosmic Visions: New Ideas in Dark
  Matter}}, 7, 2017, \href{https://arxiv.org/abs/1707.04591}{{\ttfamily
  1707.04591}}.

\bibitem{Essig:2022dfa}
R.~Essig, G.~K. Giovanetti, N.~Kurinsky, D.~McKinsey, K.~Ramanathan, K.~Stifter
  et~al., \emph{{Snowmass2021 Cosmic Frontier: The landscape of low-threshold
  dark matter direct detection in the next decade}},  in \emph{{2022 Snowmass
  Summer Study}}, 3, 2022, \href{https://arxiv.org/abs/2203.08297}{{\ttfamily
  2203.08297}}.

\bibitem{PhysRevB.103.104516}
H.~Godfrin, K.~Beauvois, A.~Sultan, E.~Krotscheck, J.~Dawidowski, B.~F\aa{}k
  et~al., \emph{Dispersion relation of landau elementary excitations and
  thermodynamic properties of superfluid $^{4}\mathrm{He}$},
  \href{https://doi.org/10.1103/PhysRevB.103.104516}{\emph{Phys. Rev. B}
  {\bfseries 103} (2021) 104516}.

\bibitem{landau1987statistical}
L.~Landau, E.~Lifshitz and L.~Pitaevskij, \emph{Statistical physics, part 2:
  theory of the condensed state}, {\emph{Course of theoretical Physics}
  {\bfseries 9} (1987) }.

\bibitem{Hertel:2018aal}
S.~A. Hertel, A.~Biekert, J.~Lin, V.~Velan and D.~N. McKinsey, \emph{{Direct
  detection of sub-GeV dark matter using a superfluid $^4$He target}},
  \href{https://doi.org/10.1103/PhysRevD.100.092007}{\emph{Phys. Rev. D}
  {\bfseries 100} (2019) 092007}
  [\href{https://arxiv.org/abs/1810.06283}{{\ttfamily 1810.06283}}].

\bibitem{Knapen:2016cue}
S.~Knapen, T.~Lin and K.~M. Zurek, \emph{{Light Dark Matter in Superfluid
  Helium: Detection with Multi-excitation Production}},
  \href{https://doi.org/10.1103/PhysRevD.95.056019}{\emph{Phys. Rev. D}
  {\bfseries 95} (2017) 056019}
  [\href{https://arxiv.org/abs/1611.06228}{{\ttfamily 1611.06228}}].

\bibitem{Schutz:2016tid}
K.~Schutz and K.~M. Zurek, \emph{{Detectability of Light Dark Matter with
  Superfluid Helium}},
  \href{https://doi.org/10.1103/PhysRevLett.117.121302}{\emph{Phys. Rev. Lett.}
  {\bfseries 117} (2016) 121302}
  [\href{https://arxiv.org/abs/1604.08206}{{\ttfamily 1604.08206}}].

\bibitem{Acanfora_2019}
F.~Acanfora, A.~Esposito and A.~D. Polosa, \emph{{Sub-GeV Dark Matter in
  Superfluid He-4: an Effective Theory Approach}},
  \href{https://doi.org/10.1140/epjc/s10052-019-7057-0}{\emph{Eur. Phys. J. C}
  {\bfseries 79} (2019) 549}
  [\href{https://arxiv.org/abs/1902.02361}{{\ttfamily 1902.02361}}].

\bibitem{Baym:2020uos}
G.~Baym, D.~H. Beck, J.~P. Filippini, C.~J. Pethick and J.~Shelton,
  \emph{{Searching for low mass dark matter via phonon creation in superfluid
  $^4$He}}, \href{https://doi.org/10.1103/PhysRevD.104.019901}{\emph{Phys. Rev.
  D} {\bfseries 102} (2020) 035014}
  [\href{https://arxiv.org/abs/2005.08824}{{\ttfamily 2005.08824}}], [Erratum:
  Phys.Rev.D 104, 019901 (2021)].

\bibitem{Caputo:2019cyg}
A.~Caputo, A.~Esposito and A.~D. Polosa, \emph{{Sub-MeV Dark Matter and the
  Goldstone Modes of Superfluid Helium}},
  \href{https://doi.org/10.1103/PhysRevD.100.116007}{\emph{Phys. Rev. D}
  {\bfseries 100} (2019) 116007}
  [\href{https://arxiv.org/abs/1907.10635}{{\ttfamily 1907.10635}}].

\bibitem{Caputo:2019xum}
A.~Caputo, A.~Esposito, E.~Geoffray, A.~D. Polosa and S.~Sun, \emph{{Dark
  Matter, Dark Photon and Superfluid He-4 from Effective Field Theory}},
  \href{https://doi.org/10.1016/j.physletb.2020.135258}{\emph{Phys. Lett. B}
  {\bfseries 802} (2020) 135258}
  [\href{https://arxiv.org/abs/1911.04511}{{\ttfamily 1911.04511}}].

\bibitem{Caputo:2019ywq}
A.~Caputo, A.~Esposito and A.~D. Polosa, \emph{{Light Dark Matter and
  Superfluid He-4 from EFT}},
  \href{https://doi.org/10.1088/1742-6596/1468/1/012060}{\emph{J. Phys. Conf.
  Ser.} {\bfseries 1468} (2020) 012060}
  [\href{https://arxiv.org/abs/1911.07867}{{\ttfamily 1911.07867}}].

\bibitem{Caputo:2020sys}
A.~Caputo, A.~Esposito, F.~Piccinini, A.~D. Polosa and G.~Rossi,
  \emph{{Directional detection of light dark matter from three-phonon events in
  superfluid $^4$He}},
  \href{https://doi.org/10.1103/PhysRevD.103.055017}{\emph{Phys. Rev. D}
  {\bfseries 103} (2021) 055017}
  [\href{https://arxiv.org/abs/2012.01432}{{\ttfamily 2012.01432}}].

\bibitem{Lanou:1987eq}
R.~E. Lanou, H.~J. Maris and G.~M. Seidel, \emph{{Detection of Solar Neutrinos
  in Superfluid Helium}},
  \href{https://doi.org/10.1103/PhysRevLett.58.2498}{\emph{Phys. Rev. Lett.}
  {\bfseries 58} (1987) 2498}.

\bibitem{Huang:2007jh}
Y.~H. Huang, R.~E. Lanou, H.~J. Maris, G.~M. Seidel, B.~Sethumadhavan and
  W.~Yao, \emph{{Potential for Precision Measurement of Solar Neutrino
  Luminosity by HERON}},
  \href{https://doi.org/10.1016/j.astropartphys.2008.06.003}{\emph{Astropart.
  Phys.} {\bfseries 30} (2008) 1}
  [\href{https://arxiv.org/abs/0711.4095}{{\ttfamily 0711.4095}}].

\bibitem{bandler1993projected}
S.~Bandler, C.~Enss, G.~Goldhaber, R.~Lanou, H.~Maris, T.~More et~al.,
  \emph{Projected performance of a large superfluid helium solar neutrino
  detector},
  \href{https://doi.org/https://doi.org/10.1007/BF00693513}{\emph{Journal of
  Low Temperature Physics} {\bfseries 93} (1993) 785}.

\bibitem{Bradley:1996cu}
D.~I. Bradley, M.~R. Follows, W.~M. Hayes and T.~Sloan, \emph{{The detection of
  low-energy neutrons and gamma-rays by superfluid He-3: A potential dark
  matter detector}},
  \href{https://doi.org/10.1016/0168-9002(95)01066-1}{\emph{Nucl. Instrum.
  Meth. A} {\bfseries 370} (1996) 141}.

\bibitem{Winkelmann:2006pw}
C.~B. Winkelmann, J.~Elbs, Y.~M. Bunkov, E.~Collin, H.~Godfrin and M.~Krusius,
  \emph{{Bolometric calibration of a superfluid He-3 detector for dark matter
  search: Direct measurement of the scintillated energy fraction for neutron,
  electron and muon events}},
  \href{https://doi.org/10.1016/j.nima.2007.01.180}{\emph{Nucl. Instrum. Meth.
  A} {\bfseries 574} (2007) 264}
  [\href{https://arxiv.org/abs/physics/0611273}{{\ttfamily physics/0611273}}].

\bibitem{Winkelmann:2006rg}
C.~B. Winkelmann, J.~Elbs, E.~Collin, Y.~M. Bunkov and H.~Godfrin,
  \emph{{ULTIMA: A bolometric detector for dark matter search using superfluid
  He-3}}, \href{https://doi.org/10.1016/j.nima.2005.12.016}{\emph{Nucl.
  Instrum. Meth. A} {\bfseries 559} (2006) 384}.

\bibitem{Lanou:1988iq}
R.~E. Lanou, H.~J. Maris and G.~M. Seidel, \emph{{SUPERFLUID HELIUM AS A DARK
  MATTER DETECTOR}},  in \emph{{23rd Rencontres de Moriond: Astronomy:
  Detection of Dark Matter}}, 1988.

\bibitem{Bandler:1991ep}
S.~R. Bandler, R.~E. Lanou, H.~J. Maris, T.~More, F.~S. Porter, G.~M. Seidel
  et~al., \emph{{Particle detection using superfluid helium}},  in \emph{{4th
  International Workshop on Low Temperature Detectors for Neutrinos and Dark
  Matter IV}}, 9, 1991.

\bibitem{Bandler:1992zz}
S.~R. Bandler, R.~E. Lanou, H.~J. Maris, T.~More, F.~S. Porter, G.~M. Seidel
  et~al., \emph{{Particle detection by evaporation from superfluid helium}},
  \href{https://doi.org/10.1103/PhysRevLett.68.2429}{\emph{Phys. Rev. Lett.}
  {\bfseries 68} (1992) 2429}.

\bibitem{Adams:1996ge}
J.~S. Adams, S.~R. Bandler, S.~M. Brouer, C.~Enss, R.~E. Lanou, H.~J. Maris
  et~al., \emph{{'HERON' as a dark matter detector?}},  in \emph{{31st
  Rencontres de Moriond: Dark Matter and Cosmology, Quantum Measurements and
  Experimental Gravitation}}, pp.~131--136, 1996.

\bibitem{Bottino:2002ry}
A.~Bottino, N.~Fornengo and S.~Scopel, \emph{{Light relic neutralinos}},
  \href{https://doi.org/10.1103/PhysRevD.67.063519}{\emph{Phys. Rev. D}
  {\bfseries 67} (2003) 063519}
  [\href{https://arxiv.org/abs/hep-ph/0212379}{{\ttfamily hep-ph/0212379}}].

\bibitem{Bottino:2003cz}
A.~Bottino, F.~Donato, N.~Fornengo and S.~Scopel, \emph{{Light neutralinos and
  WIMP direct searches}},
  \href{https://doi.org/10.1103/PhysRevD.69.037302}{\emph{Phys. Rev. D}
  {\bfseries 69} (2004) 037302}
  [\href{https://arxiv.org/abs/hep-ph/0307303}{{\ttfamily hep-ph/0307303}}].

\bibitem{Shelton:2010ta}
J.~Shelton and K.~M. Zurek, \emph{{Darkogenesis: A baryon asymmetry from the
  dark matter sector}},
  \href{https://doi.org/10.1103/PhysRevD.82.123512}{\emph{Phys. Rev. D}
  {\bfseries 82} (2010) 123512}
  [\href{https://arxiv.org/abs/1008.1997}{{\ttfamily 1008.1997}}].

\bibitem{Feng:2008ya}
J.~L. Feng and J.~Kumar, \emph{{The WIMPless Miracle: Dark-Matter Particles
  without Weak-Scale Masses or Weak Interactions}},
  \href{https://doi.org/10.1103/PhysRevLett.101.231301}{\emph{Phys. Rev. Lett.}
  {\bfseries 101} (2008) 231301}
  [\href{https://arxiv.org/abs/0803.4196}{{\ttfamily 0803.4196}}].

\bibitem{Foot:2008nw}
R.~Foot, \emph{{Mirror dark matter and the new DAMA/LIBRA results: A Simple
  explanation for a beautiful experiment}},
  \href{https://doi.org/10.1103/PhysRevD.78.043529}{\emph{Phys. Rev. D}
  {\bfseries 78} (2008) 043529}
  [\href{https://arxiv.org/abs/0804.4518}{{\ttfamily 0804.4518}}].

\bibitem{CMS:2012lmn}
{\scshape CMS} collaboration, S.~Chatrchyan et~al., \emph{{Search for Dark
  Matter and Large Extra Dimensions in pp Collisions Yielding a Photon and
  Missing Transverse Energy}},
  \href{https://doi.org/10.1103/PhysRevLett.108.261803}{\emph{Phys. Rev. Lett.}
  {\bfseries 108} (2012) 261803}
  [\href{https://arxiv.org/abs/1204.0821}{{\ttfamily 1204.0821}}].

\bibitem{Fermi-LAT:2011vow}
{\scshape Fermi-LAT} collaboration, M.~Ackermann et~al., \emph{{Constraining
  Dark Matter Models from a Combined Analysis of Milky Way Satellites with the
  Fermi Large Area Telescope}},
  \href{https://doi.org/10.1103/PhysRevLett.107.241302}{\emph{Phys. Rev. Lett.}
  {\bfseries 107} (2011) 241302}
  [\href{https://arxiv.org/abs/1108.3546}{{\ttfamily 1108.3546}}].

\bibitem{Guo:2013dt}
W.~Guo and D.~N. McKinsey, \emph{{Concept for a dark matter detector using
  liquid helium-4}},
  \href{https://doi.org/10.1103/PhysRevD.87.115001}{\emph{Phys. Rev. D}
  {\bfseries 87} (2013) 115001}
  [\href{https://arxiv.org/abs/1302.0534}{{\ttfamily 1302.0534}}].

\bibitem{Ito:2013cqa}
T.~M. Ito and G.~M. Seidel, \emph{{Scintillation of Liquid Helium for
  Low-Energy Nuclear Recoils}},
  \href{https://doi.org/10.1103/PhysRevC.88.025805}{\emph{Phys. Rev. C}
  {\bfseries 88} (2013) 025805}
  [\href{https://arxiv.org/abs/1303.3858}{{\ttfamily 1303.3858}}].

\bibitem{Osterman:2020xkb}
D.~Osterman, H.~Maris, G.~Seidel and D.~Stein, \emph{{Development of a Dark
  Matter Detector that Uses Liquid He and Field Ionization}},
  \href{https://doi.org/10.1088/1742-6596/1468/1/012071}{\emph{J. Phys. Conf.
  Ser.} {\bfseries 1468} (2020) 012071}.

\bibitem{Matchev:2021fuw}
K.~T. Matchev, J.~Smolinsky, W.~Xue and Y.~You, \emph{{Superfluid Effective
  Field Theory for dark matter direct detection}},
  \href{https://doi.org/https://doi.org/10.1007/JHEP05(2022)034}{\emph{Journal
  of High Energy Physics} {\bfseries 34} (2022) }
  [\href{https://arxiv.org/abs/2108.07275}{{\ttfamily 2108.07275}}].

\bibitem{Campbell-Deem:2022fqm}
B.~Campbell-Deem, S.~Knapen, T.~Lin and E.~Villarama, \emph{{Dark matter direct
  detection from the single phonon to the nuclear recoil regime}},
  \href{https://arxiv.org/abs/2205.02250}{{\ttfamily 2205.02250}}.

\bibitem{Kavanagh:2016xfi}
B.~J. Kavanagh and C.~A.~J. O'Hare, \emph{{Reconstructing the three-dimensional
  local dark matter velocity distribution}},
  \href{https://doi.org/10.1103/PhysRevD.94.123009}{\emph{Phys. Rev. D}
  {\bfseries 94} (2016) 123009}
  [\href{https://arxiv.org/abs/1609.08630}{{\ttfamily 1609.08630}}].

\bibitem{Lee:2012pf}
S.~K. Lee and A.~H.~G. Peter, \emph{{Probing the Local Velocity Distribution of
  WIMP Dark Matter with Directional Detectors}},
  \href{https://doi.org/10.1088/1475-7516/2012/04/029}{\emph{JCAP} {\bfseries
  04} (2012) 029} [\href{https://arxiv.org/abs/1202.5035}{{\ttfamily
  1202.5035}}].

\bibitem{Radick:2020qip}
A.~Radick, A.-M. Taki and T.-T. Yu, \emph{{Dependence of Dark Matter - Electron
  Scattering on the Galactic Dark Matter Velocity Distribution}},
  \href{https://doi.org/10.1088/1475-7516/2021/02/004}{\emph{JCAP} {\bfseries
  02} (2021) 004} [\href{https://arxiv.org/abs/2011.02493}{{\ttfamily
  2011.02493}}].

\bibitem{Necib:2018igl}
L.~Necib, M.~Lisanti, S.~Garrison-Kimmel, A.~Wetzel, R.~Sanderson, P.~F.
  Hopkins et~al., \emph{{Under the Firelight: Stellar Tracers of the Local Dark
  Matter Velocity Distribution in the Milky Way}},
  \href{https://arxiv.org/abs/1810.12301}{{\ttfamily 1810.12301}}.

\bibitem{Ibarra:2017mzt}
A.~Ibarra and A.~Rappelt, \emph{{Optimized velocity distributions for direct
  dark matter detection}},
  \href{https://doi.org/10.1088/1475-7516/2017/08/039}{\emph{JCAP} {\bfseries
  08} (2017) 039} [\href{https://arxiv.org/abs/1703.09168}{{\ttfamily
  1703.09168}}].

\bibitem{Silver:1989zz}
R.~N. Silver, \emph{{Theory of deep inelastic neutron scattering. 2.
  Application to normal and superfluid He-4}},
  \href{https://doi.org/10.1103/PhysRevB.39.4022}{\emph{Phys. Rev. B}
  {\bfseries 39} (1989) 4022}.

\bibitem{griffin1993excitations}
A.~Griffin, \emph{Excitations in a Bose-condensed Liquid}, Cambridge Studies in
  Low Temperature Physics. Cambridge University Press, 1993,
  \href{https://doi.org/10.1017/CBO9780511524257}{10.1017/CBO9780511524257}.

\bibitem{campbell2015dynamic}
C.~Campbell, E.~Krotscheck and T.~Lichtenegger, \emph{Dynamic many-body theory:
  Multiparticle fluctuations and the dynamic structure of he 4},
  {\emph{Physical Review B} {\bfseries 91} (2015) 184510}.

\bibitem{landau1941theory}
L.~Landau, \emph{Theory of the superfluidity of helium ii},
  \href{https://doi.org/10.1103/PhysRev.60.356}{\emph{Phys. Rev.} {\bfseries
  60} (1941) 356}.

\bibitem{Nicolis_2018}
A.~Nicolis and R.~Penco, \emph{{Mutual Interactions of Phonons, Rotons, and
  Gravity}}, \href{https://doi.org/10.1103/PhysRevB.97.134516}{\emph{Phys. Rev.
  B} {\bfseries 97} (2018) 134516}
  [\href{https://arxiv.org/abs/1705.08914}{{\ttfamily 1705.08914}}].

\bibitem{PhysRevLett.119.260402}
Y.~Castin, A.~Sinatra and H.~Kurkjian, \emph{Landau phonon-roton theory
  revisited for superfluid $^{4}\mathrm{He}$ and fermi gases},
  \href{https://doi.org/10.1103/PhysRevLett.119.260402}{\emph{Phys. Rev. Lett.}
  {\bfseries 119} (2017) 260402}.

\bibitem{PhysRev.38.1342}
P.~Rudnick, \emph{The capture and loss of electrons by helium ions in helium},
  \href{https://doi.org/10.1103/PhysRev.38.1342}{\emph{Phys. Rev.} {\bfseries
  38} (1931) 1342}.

\bibitem{PhysRevA.9.2434}
P.~Hvelplund and E.~H. Pedersen, \emph{Single and double electron loss by fast
  helium atoms in gases},
  \href{https://doi.org/10.1103/PhysRevA.9.2434}{\emph{Phys. Rev. A} {\bfseries
  9} (1974) 2434}.

\bibitem{cramer1957elastic}
W.~Cramer and J.~Simons, \emph{Elastic and inelastic scattering of low-velocity
  he+ ions in helium}, {\emph{The Journal of chemical physics} {\bfseries 26}
  (1957) 1272}.

\bibitem{Hegerberg_1978}
R.~Hegerberg, T.~Stefansson and M.~T. Elford, \emph{Measurement of the
  symmetric charge-exchange cross section in helium and argon in the impact
  energy range 1-10 {keV}},
  \href{https://doi.org/10.1088/0022-3700/11/1/017}{\emph{Journal of Physics B:
  Atomic and Molecular Physics} {\bfseries 11} (1978) 133}.

\bibitem{PhysRevA.39.4440}
R.~D. DuBois, \emph{Multiple ionization in ${\mathrm{he}}^{+}$--rare-gas
  collisions}, \href{https://doi.org/10.1103/PhysRevA.39.4440}{\emph{Phys. Rev.
  A} {\bfseries 39} (1989) 4440}.

\bibitem{pivovar1962electron}
L.~Pivovar, V.~Tabuev and M.~Novikov, \emph{Electron loss and capture by
  200--1500 kev helium ions in various gases}, {\emph{Sov. Phys. JETP}
  {\bfseries 14} (1962) 20}.

\bibitem{PhysRevA.32.829}
M.~E. Rudd, T.~V. Goffe, A.~Itoh and R.~D. DuBois, \emph{Cross sections for
  ionization of gases by 10--2000-kev ${\mathrm{he}}^{+}$ ions and for electron
  capture and loss by 5--350-kev ${\mathrm{he}}^{+}$ ions},
  \href{https://doi.org/10.1103/PhysRevA.32.829}{\emph{Phys. Rev. A} {\bfseries
  32} (1985) 829}.

\bibitem{PhysRevA.20.1816}
R.~D. Rivarola and R.~D. Piacentini, \emph{Differential and total cross
  sections in ${\mathrm{he}}^{2+}$-he collisions},
  \href{https://doi.org/10.1103/PhysRevA.20.1816}{\emph{Phys. Rev. A}
  {\bfseries 20} (1979) 1816}.

\bibitem{Grozdanov_1980}
T.~P. Grozdanov and R.~K. Janev, \emph{Two-electron capture in slow ion-atom
  collisions}, \href{https://doi.org/10.1088/0022-3700/13/17/021}{\emph{Journal
  of Physics B: Atomic and Molecular Physics} {\bfseries 13} (1980) 3431}.

\bibitem{Fulton_1966}
M.~J. Fulton and M.~H. Mittleman, \emph{Electron capture by helium ions in
  neutral helium},
  \href{https://doi.org/10.1088/0370-1328/87/3/307}{\emph{Proceedings of the
  Physical Society} {\bfseries 87} (1966) 669}.

\bibitem{hertel1964cross}
G.~Hertel and W.~Koski, \emph{Cross sections for the single charge transfer of
  doubly-charged rare-gas ions in their own gases}, {\emph{The Journal of
  Chemical Physics} {\bfseries 40} (1964) 3452}.

\bibitem{Stich_1985}
W.~Stich, H.~J. Ludde and R.~M. Dreizler, \emph{{TDHF} calculations for
  two-electron systems},
  \href{https://doi.org/10.1088/0022-3700/18/6/019}{\emph{Journal of Physics B:
  Atomic and Molecular Physics} {\bfseries 18} (1985) 1195}.

\bibitem{RevModPhys.30.1137}
S.~K. Allison, \emph{Experimental results on charge-changing collisions of
  hydrogen and helium atoms and ions at kinetic energies above 0.2 kev},
  \href{https://doi.org/10.1103/RevModPhys.30.1137}{\emph{Rev. Mod. Phys.}
  {\bfseries 30} (1958) 1137}.

\bibitem{WU198857}
W.~Wu, B.~Huber and K.~Wiesemann, \emph{Cross sections for electron capture by
  neutral and charged particles in collisions with he},
  \href{https://doi.org/https://doi.org/10.1016/0092-640X(88)90004-6}{\emph{Atomic
  Data and Nuclear Data Tables} {\bfseries 40} (1988) 57}.

\bibitem{Hvelplund_1976}
P.~Hvelplund, J.~Heinemeier, E.~H. Pedersen and F.~R. Simpson, \emph{Electron
  capture by fast he2$+$ions in gases},
  \href{https://doi.org/10.1088/0022-3700/9/3/017}{\emph{Journal of Physics B:
  Atomic and Molecular Physics} {\bfseries 9} (1976) 491}.

\bibitem{doi:10.1246/bcsj.49.933}
S.~Sato, K.-i. Kowari and K.~Okazaki, \emph{A comparison of the primary
  processes of the radiolyses induced by $\alpha$-, $\beta$-, and
  $\gamma$-rays}, \href{https://doi.org/10.1246/bcsj.49.933}{\emph{Bulletin of
  the Chemical Society of Japan} {\bfseries 49} (1976) 933}
  [\href{https://arxiv.org/abs/https://doi.org/10.1246/bcsj.49.933}{{\ttfamily
  https://doi.org/10.1246/bcsj.49.933}}].

\bibitem{thomas_1927}
L.~H. Thomas, \emph{The production of characteristic x-rays by electronic
  impact}, \href{https://doi.org/10.1017/S0305004100015620}{\emph{Mathematical
  Proceedings of the Cambridge Philosophical Society} {\bfseries 23} (1927)
  829–831}.

\bibitem{PhysRev.124.128}
J.~Lindhard and M.~Scharff, \emph{Energy dissipation by ions in the kev
  region}, \href{https://doi.org/10.1103/PhysRev.124.128}{\emph{Phys. Rev.}
  {\bfseries 124} (1961) 128}.

\bibitem{PhysRev.135.A1575}
H.~C. Hayden and N.~G. Utterback, \emph{Ionization of helium, neon, and
  nitrogen by helium atoms},
  \href{https://doi.org/10.1103/PhysRev.135.A1575}{\emph{Phys. Rev.} {\bfseries
  135} (1964) A1575}.

\bibitem{PhysRev.109.355}
C.~F. Barnett and H.~K. Reynolds, \emph{Charge exchange cross sections of
  hydrogen particles in gases at high energies},
  \href{https://doi.org/10.1103/PhysRev.109.355}{\emph{Phys. Rev.} {\bfseries
  109} (1958) 355}.

\bibitem{PhysRev.178.271}
L.~J. Puckett, G.~O. Taylor and D.~W. Martin, \emph{Cross sections for ion and
  electron production in gases by 0.15-1.00-mev hydrogen and helium ions and
  atoms}, \href{https://doi.org/10.1103/PhysRev.178.271}{\emph{Phys. Rev.}
  {\bfseries 178} (1969) 271}.

\bibitem{PhysRevA.36.2585}
R.~D. DuBois, \emph{Ionization and charge transfer in
  ${\mathrm{he}}^{2+}$--rare-gas collisions. ii},
  \href{https://doi.org/10.1103/PhysRevA.36.2585}{\emph{Phys. Rev. A}
  {\bfseries 36} (1987) 2585}.

\bibitem{PhysRevA.63.062717}
A.~C.~F. Santos, W.~S. Melo, M.~M. Sant'Anna, G.~M. Sigaud and E.~C.
  Montenegro, \emph{Absolute multiple-ionization cross sections of noble gases
  by ${\mathrm{he}}^{+}$},
  \href{https://doi.org/10.1103/PhysRevA.63.062717}{\emph{Phys. Rev. A}
  {\bfseries 63} (2001) 062717}.

\bibitem{Shah_1985}
M.~B. Shah and H.~B. Gilbody, \emph{Single and double ionisation of helium by
  h$+$, he2$+$and li3$+$ions},
  \href{https://doi.org/10.1088/0022-3700/18/5/010}{\emph{Journal of Physics B:
  Atomic and Molecular Physics} {\bfseries 18} (1985) 899}.

\bibitem{osti_4196582}
V.~Kempter, F.~Veith and L.~Zehnle, \emph{Excitation processes-in low-energy
  collisions between ground state helium atoms},
  \href{https://doi.org/10.1088/0022-3700/8/7/010}{\emph{J. Phys., B (London),
  v. 8, no. 7, pp. 1041-1052} {\bfseries 8} (1975) }.

\bibitem{Heer1965ExcitationOH}
F.~J. de~Heer and J.~van~den Bos, \emph{Excitation of helium by he + and
  polarization of the resulting radiation}, {\emph{Physica D: Nonlinear
  Phenomena} {\bfseries 31} (1965) 365}.

\bibitem{Mei:2007jn}
D.~M. Mei, Z.~B. Yin, L.~C. Stonehill and A.~Hime, \emph{{A Model of Nuclear
  Recoil Scintillation Efficiency in Noble Liquids}},
  \href{https://doi.org/10.1016/j.astropartphys.2008.06.001}{\emph{Astropart.
  Phys.} {\bfseries 30} (2008) 12}
  [\href{https://arxiv.org/abs/0712.2470}{{\ttfamily 0712.2470}}].

\bibitem{Ito:2011cy}
T.~M. Ito, S.~M. Clayton, J.~Ramsey, M.~Karcz, C.~Y. Liu, J.~C. Long et~al.,
  \emph{{Effect of an electric field on superfluid helium scintillation
  produced by $\alpha$-particle sources}},
  \href{https://doi.org/10.1103/PhysRevA.85.042718}{\emph{Phys. Rev. A}
  {\bfseries 85} (2012) 042718}
  [\href{https://arxiv.org/abs/1110.0570}{{\ttfamily 1110.0570}}].

\bibitem{Adams:1995mk}
J.~S. Adams, S.~R. Bandler, S.~M. Brouer, R.~E. Lanou, H.~J. Maris, T.~More
  et~al., \emph{{Simultaneous calorimetric detection of rotons and photons
  generated by particles in superfluid helium}},
  \href{https://doi.org/10.1016/0370-2693(94)01423-A}{\emph{Phys. Lett. B}
  {\bfseries 341} (1995) 431}.

\bibitem{bennewitz1972he}
H.~Bennewitz, H.~Busse, H.~Dohmann, D.~Oates and W.~Schrader, \emph{He 4-he 4
  interaction potential from low energy total cross section measurements},
  \href{https://doi.org/https://doi.org/10.1007/BF01379683}{\emph{Zeitschrift
  f{\"u}r Physik A Hadrons and nuclei} {\bfseries 253} (1972) 435}.

\bibitem{feltgen1973determination}
R.~Feltgen, H.~Pauly, F.~Torello and H.~Vehmeyer, \emph{Determination of the
  ${\mathrm{he}}^{4}$-${\mathrm{he}}^{4}$ repulsive potential up to 0.14 ev by
  inversion of high-resolution total---cross-section measurements},
  \href{https://doi.org/10.1103/PhysRevLett.30.820}{\emph{Phys. Rev. Lett.}
  {\bfseries 30} (1973) 820}.

\bibitem{bishop1977low}
R.~Bishop, H.~Ghassib and M.~Strayer, \emph{Low-energy he-he interactions with
  phenomenological potentials},
  \href{https://doi.org/https://doi.org/10.1007/BF00654874}{\emph{Journal of
  Low Temperature Physics} {\bfseries 26} (1977) 669}.

\bibitem{calogero1967variable}
F.~Calogero, \emph{Variable phase approach to potential scattering},
  \href{https://doi.org/https://doi.org/10.1119/1.1975005}{\emph{American
  Journal of Physics} {\bfseries 36} (1968) 566}.

\bibitem{miller1969wkb}
W.~H. Miller, \emph{Wkb solution of inversion problems for potential
  scattering},
  \href{https://doi.org/https://doi.org/10.1063/1.1672572}{\emph{The Journal of
  Chemical Physics} {\bfseries 51} (1969) 3631}.

\bibitem{miller1971additional}
W.~H. Miller, \emph{Additional wkb inversion relations for bound-state and
  scattering problems},
  \href{https://doi.org/https://doi.org/10.1063/1.1674655}{\emph{The Journal of
  Chemical Physics} {\bfseries 54} (1971) 4174}.

\bibitem{silver1988theory}
R.~N. Silver, \emph{Theory of deep inelastic neutron scattering on quantum
  fluids}, {\emph{Physical Review B} {\bfseries 37} (1988) 3794}.

\bibitem{landau1949theory}
L.~Landau and I.~Khalatnikov, \emph{Theory of viscosity of hellium ii. 1.
  collissions of elementary excitations in hellium ii.},
  \href{https://doi.org/https://doi.org/10.1016/B978-0-08-010586-4.50074-2}{\emph{Zhurnal
  Eksperimentalnoi i Teoreticheskoi Fiziki} {\bfseries 19} (1949) 637}.

\bibitem{1965511}
L.~Landau, \emph{70 - the theory of the viscosity of helium ii: Ii. calculation
  of the viscosity coefficient},  in \emph{Collected Papers of L.D. Landau}
  (D.~{TER HAAR}, ed.), pp.~511--531.
\newblock Pergamon, 1965.
\newblock
  \href{https://doi.org/https://doi.org/10.1016/B978-0-08-010586-4.50075-4}{DOI}.

\bibitem{Dodelson:2003ft}
S.~Dodelson, \emph{{Modern Cosmology}}. Academic Press, Amsterdam, 2003.

\bibitem{Weinberg:2008zzc}
S.~Weinberg, \emph{{Cosmology}}. Oxford University Press, 2008.

\bibitem{donnelly1981specific}
R.~Donnelly, J.~Donnelly and R.~Hills, \emph{Specific heat and dispersion curve
  for helium ii}, {\emph{Journal of Low Temperature Physics} {\bfseries 44}
  (1981) 471}.

\bibitem{MSXY}
K.~Gunther, Y.~Lee, K.~Matchev, T.~Saab, J.~Smolinsky, W.~Xue et~al., in
  preparation.

\bibitem{Feldman:1997qc}
G.~J. Feldman and R.~D. Cousins, \emph{{A Unified approach to the classical
  statistical analysis of small signals}},
  \href{https://doi.org/10.1103/PhysRevD.57.3873}{\emph{Phys. Rev. D}
  {\bfseries 57} (1998) 3873}
  [\href{https://arxiv.org/abs/physics/9711021}{{\ttfamily physics/9711021}}].

\bibitem{Lista:2016chp}
L.~Lista, \emph{{Practical Statistics for Particle Physicists}},  in
  \emph{{2016 European School of High-Energy Physics}}, pp.~213--258, 2017,
  \href{https://arxiv.org/abs/1609.04150}{{\ttfamily 1609.04150}},
  \href{https://doi.org/10.23730/CYRSP-2017-005.213}{DOI}.

\bibitem{Bhat:2010zqs}
P.~C. Bhat, \emph{{Advanced Analysis Methods in Particle Physics}}, .

\bibitem{Sinervo:2002sa}
P.~K. Sinervo, \emph{{Signal significance in particle physics}},  in
  \emph{{Conference on Advanced Statistical Techniques in Particle Physics}},
  pp.~64--76, 6, 2002, \href{https://arxiv.org/abs/hep-ex/0208005}{{\ttfamily
  hep-ex/0208005}}.

\bibitem{Barlow:2003cx}
R.~Barlow, \emph{{Introduction to statistical issues in particle physics}},
  {\emph{eConf} {\bfseries C030908} (2003) MOAT002}
  [\href{https://arxiv.org/abs/physics/0311105}{{\ttfamily physics/0311105}}].

\end{thebibliography}\endgroup

\end{document}